\documentclass{aa}  

\usepackage{graphicx}
\usepackage{txfonts}
\usepackage{pdflscape}	
\usepackage{color, colortbl}
\usepackage[colorlinks, citecolor=blue, linkcolor=blue, urlcolor=blue]{hyperref}
\begin{document}

   \title{Humps and bumps: The effects of shocks on the optical light curves of fundamental-mode RR Lyrae stars}


   \author{Z. Prudil\inst{1}
          \and
          I. D\'ek\'any\inst{1}
          \and
          R. Smolec\inst{2}
          \and
          M. Catelan\inst{3,4}
          \and
          E. K. Grebel\inst{1}
          \and
          A. Kunder\inst{5}}

\institute{Astronomisches Rechen-Institut, Zentrum f{\"u}r Astronomie der Universit{\"a}t Heidelberg, M{\"o}nchhofstr. 12-14, D-69120 Heidelberg, Germany \\ \email{prudilz@ari.uni-heidelberg.de}
		\and
		Nicolaus Copernicus Astronomical Center, Polish Academy of Sciences, ul. Bartycka 18, 00-716 Warszawa, Poland
		\and
		Instituto de Astrof{\'i}sica, Pontificia Universidad Cat{\'o}lica de Chile, Av. Vicu{\~n}a Mackenna 4860, 7820436 Macul, Santiago, Chile
		\and
		Instituto Milenio de Astrof{\'i}sica, Av. Vicu{\~n}a Mackenna 4860, 7820436 Macul, Santiago, Chile
		\and 
		Saint Martin's University, 5000 Abbey Way SE, Lacey, WA, 98503}

   \date{} 

  \abstract
	{We present the most extended and homogeneous study carried out so far of the main and early shocks in 1485 RR~Lyrae stars in the Galactic bulge observed by the Optical Gravitational Lensing Experiment (OGLE). We selected non-modulated fundamental-mode RR~Lyrae stars with good-quality photometry. Using a self-developed method, we determined the centers and strengths of main and early shock features in the phased light curves. We found that the position of both humps and bumps are highly correlated with the pulsation properties of the studied variables. Pulsators with a pronounced main shock are concentrated in the low-amplitude regime of the period-amplitude diagram, while stars with a strong early shock have average and above-average pulsation amplitudes. A connection between the main and early shocks and the Fourier coefficients is also observed. In the color-magnitude diagram (CMD), we see a separation between stars with strong and weak shocks. Variables with a pronounced main shock cluster close to the fundamental red edge of the instability strip (IS), while stars with a strong early shock tend to clump in the center and near the fundamental blue edge of the IS. The appearance of shocks and their properties seem independent of the direction of evolution estimated from the period change rate of the studied stars. In addition, the differences in the period change rate between the two main Oosterhoff groups found in the Galactic bulge suggest that stars of Oosterhoff type I are located close to the zero-age horizontal branch while Oosterhoff type II variables are on their way toward the fundamental red edge of the instability strip, thus having already left the zero-age horizontal branch.}

\titlerunning{The effects of shocks on RR~Lyrae light curves}
 
\maketitle


\section{Introduction}

RR~Lyrae stars are dominantly radial pulsators. These horizontal branch stars are tracers of the old population. They are utilized in various fields of astronomy, mainly in studies focused on stellar pulsations \citep[e.g. ][]{Szabo2010,Kollath2011,Smolec2015}, on the Milky Way (MW) subcomponents such as the Galactic disk \citep[e.g. ][]{Dekany2018,Mateu2018}, the Galactic bulge \citep[e.g.][]{Dekany2013,Pietrukowicz2015}, and the Galactic halo \citep[e.g. ][]{Drake2013,Belokurov2018,Hernitschek2018}, globular clusters \citep[e.g. ][]{Cacciari2005,Catelan2009}, and the local extragalactic neighborhood such as the Magellanic Clouds \citep[e.g. ][]{Haschke2012RRLyrLMC,Haschke2012RRLyrSMC,JD2017RRlyrae} and neighboring dwarf galaxies \citep[e.g.][]{Pietrzynski2008,Karczmarek2015,Karczmarek2017}. They are divided into three categories based on their radial pulsation mode, namely RRab -- fundamental mode, RRc -- first overtone and RRd -- double-mode RR~Lyrae stars, which are simultaneously pulsating in the first overtone and fundamental mode \citep[see, e.g.,][]{Catelan2015book}.

Despite their plentitude in the MW \citep[over 100\,000 stars in Gaia DR2;][]{Clementini2018}; several open issues hamper their full potential as e.g. stellar tracers. Some examples include the Blazhko effect \citep{Blazhko1907}, causing a modulation of the light curves, and the scarcity of RR~Lyrae stars in binary systems \citep{Hajdu2015,Liska2016b,Prudil2019Binary} and therefore the absence of accurate mass determinations. Also, the origin of the Oosterhoff dichotomy \citep{Oosterhoff1939}, which divides globular clusters that contain RR~Lyrae variables into two groups based on their pulsation properties, is far from being resolved. Several possible solutions emerged over the past decades, e.g., increased helium abundance for RR~Lyrae stars in Oosterhoff type\.II (Oo\,II) clusters \citep{Sandage1981KS,Sandage1981} or a different direction in crossing the instability strip \citep{vanAlbada1973}. In more recent years some progress has been made suggesting that population and/or metallicity effects might be the cause of the Oosterhoff dichotomy, e.g., \citet{Catelan2009,Lee2014,Fabrizio2019}, especially among field RR~Lyrae variables.

In this study, we focus on shock phenomena in RR~Lyrae stars. In their pioneering work, \citet{Struve1948} observed line doubling and line emissions of hydrogen in the spectrum of the prototype of RR~Lyrae stars, RR~Lyr itself, during certain pulsation phases. Further spectroscopic studies \citep[e.g.][]{Preston1964,Preston1965} confirmed the findings of \citet{Struve1948} and added more features that appear at certain phases of a pulsation cycle. Today we associate these aspects with shock waves propagating through the atmospheres of RR~Lyrae stars. The Balmer lines forming in the upper layers of atmospheres are mainly affected \citep{Hill1972,Gillet1988}, although some of the metallic lines are distorted through line broadening and absorption line doubling \citep{Chadid1996doub,Chadid1996broa}. Recently, \cite{Preston2009} reported helium emission and absorption lines observed during the shock events in some of the RR~Lyrae stars. Also, \citet{Chadid2017} showed that in the upper atmosphere the main shock intensity is larger among metal-poor RR~Lyrae stars than in metal-rich RR~Lyrae variables.

These shock waves correlate with features of the RRab light curves known as the hump and bump \citep{Christy1966}. The shock waves are a consequence of rapid compression events in the atmospheres of RR Lyrae stars that oscillate with large amplitudes due to the opacity mechanism. The hump is a result of a sudden stop of the infalling photosphere and immediate outward expansion, while the bump is connected with the collision of the infalling material before reaching the minimum radius \citep[e.g.,][]{Hill1972,Fokin1992}.

In this paper, we study the impact of humps and bumps on the phased light curves of non-modulated fundamental mode RR~Lyrae stars. In Sec.~\ref{sec:CritStudStars} we describe the selection criteria for the studied sample. Section \ref{sec:AnalysStrengh} outlines the adopted procedure to estimate the strength of individual humps and bumps in the light curves. Sec.~\ref{sec:centersANDeffect} discusses the positions of the centers of the humps and bumps, and the effect of main and early shocks on the light curves of RR~Lyrae stars. In Sec.~\ref{sec:ShocksBigPicture} we compare how our findings relate to the stars' positions in the color-magnitude diagram and their evolutionary direction. In the last Section \ref{sec:Conclus} we summarize our results. 

\section{Sample selection} \label{sec:CritStudStars}

To study small features in the light curve shapes of RR~Lyrae stars, e.g. humps and bumps, one needs precise, homogeneous, and well-sampled photometry. Moreover, information about any alterations in the shapes of the light curves, e.g. modulation \citep{Blazhko1907}, additional modes, or possible binarity, is necessary since they would inevitably complicate the analysis and overall results.

The aforementioned criteria are fulfilled by the sample collected by the OGLE-IV survey \citep{Udalski2015} of more than 38000 RR~Lyrae stars in the Galactic bulge \citep{Soszynski2014OGLEIV,Soszynski2017OGLEIV}. OGLE-IV provides photometry in two passbands $V$ and $I$, where the latter contains substantially more measurements. The number of epochs collected for individual variables ranges from dozens up to tens of thousands. For the purpose of our study, we used the light curves of fundamental-mode RR~Lyrae stars without any sign of modulation, additional modes, or binarity \citep{Hajdu2015,Prudil2019Binary}. We used a sample of non-modulated RR~Lyrae stars collected in a study by \citet{Prudil2019OO}. In addition to the information provided by OGLE, this sample contains the following additional information for individiual objects: estimates of the metallicity based on the photometric data, absolute magnitudes, distances, amplitudes, and mean apparent magnitudes in the $K_{s}$-band from the VVV photometric survey \citep[with a crossmatch radius $\approx {0.5}''$,][]{Minniti2010VVV}. Furthermore, each star is associated with one of the two Oosterhoff groups \citep{Oosterhoff1939} in the Galactic bulge \citep{Prudil2019OO}.

To address the criteria for precision in photometry and sufficient phase coverage of the light curves, we removed stars fainter than 16.5\,mag\footnote{Stars brighter than 16.5\,mag in the $I$-band have mean photometric errors below 0.01\,mag in OGLE-IV.} and stars with less than 1000 epochs in the $I$-band. The data in the aforementioned passband served as a primary source of photometry for this study. The remaining variables that surpass the prior criteria were visually examined on the basis of their phased light curves. Stars with large scatter and/or with a high number of outliers were removed. In the end, from the original sample containing 8141 RR~Lyrae stars, 1485 variables remained, and serve as the studied sample for the remainder of this paper.

\section{Analysis of the impact of humps and bumps on the phased light curves} \label{sec:AnalysStrengh}

Our initial approach to estimate the significance of humps and bumps in the phased diagrams of RR~Lyrae stars was strictly subjective. We visually classified all RR~Lyrae phase-folded light curves into four categories (1 - 4) based on the visual prominence of the aforementioned phase curve features (with 4 standing for a strong and 1 for a weak shock effect on the phased light curve). We note that this categorization was purely informative and in the following analysis served only for the assessment of the calculated values, in order to decide which of them best describes the impact of the humps and bumps on the phased light curves (the visual classification of the humps and bumps is included in Tab.~\ref{tab:KinProp}).

In order to quantitatively estimate the strength of the humps and bumps, we developed two analogous approaches (described in following subsections) using the Fourier decomposition of the phased light curves and the Gaussian fitting of the residual. 

\begin{figure*}
\centering
\includegraphics[width=2\columnwidth]{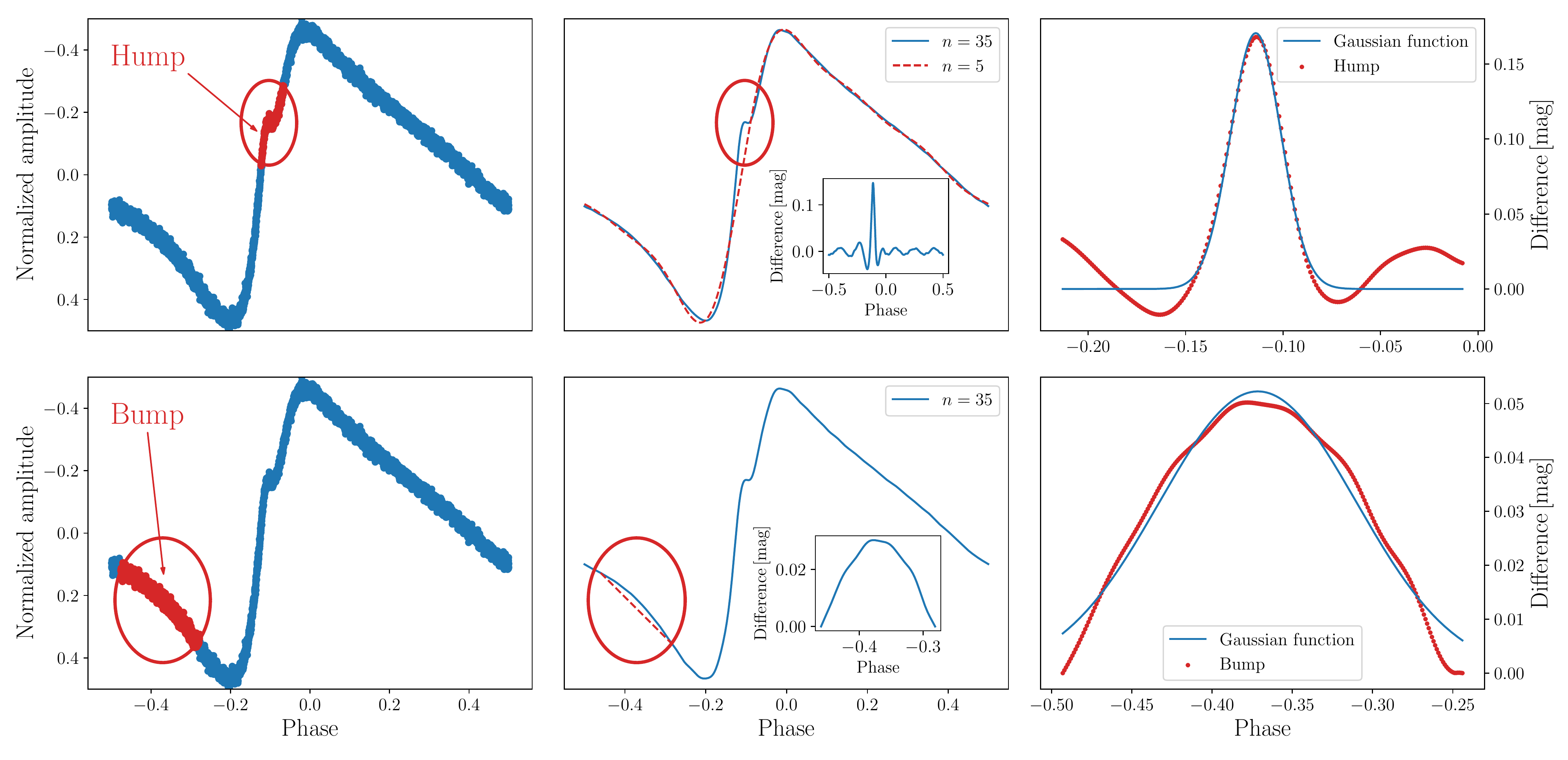}
\caption{Visual description of the analysis of the hump (top panels) and bump (bottom panels) features in one of the sample stars. The left panels depict the phased light curve of a selected star with $I$-band photometric data represented with blue dots. The red ellipse denotes the shock features: hump (top left panel) and bump (bottom left panel). The red dots denote the removed portions of the phased light curve. The middle panels show the model light curve described by a high-order Fourier ($n=35$ in the legend of the panels) fit as given by Eq.~\ref{eq:FourierSeries} (blue solid line). 
The dashed red line in the top middle panel represents a low-order Fourier fit ($n=5$ in the legend of the panel) to the phased light curve with the hump region removed (i.e., the section of the light curve encompassed by the red ellipse), and the inset in the same panel shows the difference between the high- and low-order fits. The bottom middle panel shows a linear fit (red dashed line) for the area around the bump, and the inset shows the difference between the fit of a high-order Fourier series and the linear fit in the vicinity of the bump. The right-hand panels show the result of subtracting the linear fits from the Fourier models (red dots), as fitted with a Gaussian (blue solid line).}
\label{fig:HumpCenter}
\end{figure*}

The initial steps for both approaches were similar in phasing the light curves and normalizing the magnitudes from the observed data. This way we scaled the time (Heliocentric Julian Date) and magnitude component of the data to encompass a range of $\approx\pm$0.5 for all stars and both parameters become unitless (see the top and bottom panels of Fig.~\ref{fig:HumpCenter}). 

For each phased light curve we then visually marked the approximate center of the hump and bump $\left( x_{\rm HUMP}, y_{\rm HUMP}, x_{\rm BUMP}, y_{\rm BUMP} \right)$ on the descending/ascending branch light curve sections (see red squares in the left-hand panels of Fig.~\ref{fig:HumpCenter}). We note that we center our phased light curves at zero (maximum brightness), thus all values of $x_{\rm HUMP}$ and $x_{\rm BUMP}$ are negative. The positions for humps and bumps on the phased light curves are discussed in the Sec.~\ref{sec:CentersOfHumpBump}.

\subsection{Strength of a hump}

For the study of the main shock (hump) in the atmospheres of RR~Lyrae stars, we proceeded in the following way. First, we fitted the $I$-band light curves with a truncated Fourier series:

\begin{equation} \label{eq:FourierSeries}
m\left ( t \right ) = A_{0}^{I} + \sum_{k=1}^{n} A_{k}^{I} \cdot \text{cos} \left (2\pi k \vartheta + \varphi_{k}^{I} \right ),
\end{equation} 
where the $A_{k}^{I}$ stand for amplitudes and $\varphi_{k}^{I}$ represent phases. The $\vartheta$ is a phase function, which is described as the ratio between the time difference of the Heliocentric Julian Date and time of maximum brightness $HJD-M_{0}$, and pulsation period $P$: $\vartheta=\left(HJD-M_{0}\right)/P$. In order to properly describe small features between brightness minimum and maximum, we chose for all stars a high order of the Fourier fit of $n = 35$ (blue solid line in the middle panels of Fig.~\ref{fig:HumpCenter}). This high degree thoroughly described the region between the brightness minima and maxima where the hump occurs. We emphasize that pulsation properties like total amplitude and Fourier parameters were determined based on observed, unnormalized data.

Secondly, using the provisional centers of the hump in the normalized light curves space we removed the portion of the phased light curve containing the hump (red ellipse in the top left-hand panel of Fig.~\ref{fig:HumpCenter}). The size of the removed portion was $0.15$ wide in phase, and $0.3$ high in normalized magnitude (denoted by the red ellipses in Fig.~\ref{fig:HumpCenter})\footnote{The cut out region was based on the center of the hump and bump $\left( x_{\rm HUMP}, y_{\rm HUMP}, x_{\rm BUMP}, y_{\rm BUMP} \right)$ fulfiling following criteria $x\pm0.075$ and $y\pm0.15$.}. This trimmed phased light curve was again decomposed using the low-order Fourier series (in the majority of the cases, orders from four up to fifteen were used) tailored for each phased light curve individually (red dashed line in the top middle panel of Fig.~\ref{fig:HumpCenter}). 

In the next step, we subtracted both Fourier models (red and blue lines in top middle panel of the Fig.~\ref{fig:HumpCenter}) and fitted the hump region using a Gaussian function (see top right-hand panel of Fig.~\ref{fig:HumpCenter}). We integrated the area under the Gaussian function to obtain the amplitude (height of the hump), the full width half maximum (FWHM) of the hump, and the area under the Gaussian curve. An example of a fitted Fourier difference is shown in the top right-hand panel of Fig.~\ref{fig:HumpCenter}. 

\subsection{Strength of a bump}

In the case of the early shock, i.e., the bump, we proceeded in a similar way as for the main shock. 

First, we decomposed the full phased light curve with a low degree Fourier series, since the area prior to the brightness minimum is more sensitive to overfitting. Subsequently, we removed the area around the bump, using the bump's approximate center and width in phase space $\pm$\,0.125 for each light curve separately. Since the Fourier decomposition of the light curve with a large phase gap led to unrealistic results we adopted a different approach than in the case of the hump using solely the removed region of the Fourier model. For this clipped region, we determined the line connecting the beginning and end points. Then, we subtractred this straight line from the Fourier model of the bump in the relevant phase range -- $\pm 0.125$ (see the inset in the bottom middle panel of Fig.~\ref{fig:HumpCenter}). We then fitted the resulting difference using a Gaussian function (see bottom right-hand panel of Fig.~\ref{fig:HumpCenter}) and integrated over it, obtaining, as in the case of the hump, its amplitude (height of the bump), the FWHM of the bump, and the area under the Gaussian curve. 

In the last step, we compared the visually analyzed sample (classified into four groups, see Sec.~\ref{sec:AnalysStrengh}) with the obtained Gaussian properties of the humps and bumps. We conclude that the amplitude of the Gaussian curves describes the strength of the bump/hump features best, as it correlates well with the four categories we assigned to the light curves based on the visual prominence of hump/bump features, where classes 1 and 2 (denoting a negligible hump/bump) have small amplitudes and vice versa for classes 3 and 4 where we observe a prominent distortion and high amplitudes ($r_{\rm vis}^{\rm amp} = 0.86$ and $r_{\rm vis}^{\rm amp} = 0.72$, for hump and bump, respectively). The classes for individual stars together with the amplitudes for their shock features can be found in Tab.~\ref{sec:AppTable}, and the full table is enclosed in the supplementary material. Thus, for the remainder of this study, we will use it as a qualitative estimate of the shock impact on the phased light curve. We note that the amplitudes of humps and bumps are more than three times larger than the photometric errors of the dataset, therefore all amplitude detections are significant.

\section{Humps and bumps in the phased light curves} \label{sec:centersANDeffect}

In this section, we will discuss the positions and effects of humps and bumps on the phased light curves. We use the determined centers of humps and bumps ($x_{\rm HUMP}, y_{\rm HUMP}$, $x_{\rm BUMP}, y_{\rm BUMP}$) which represent the position in the phase $x$ and in the normalized magnitudes $y$. 

\subsection{Timing of the hump and bump} \label{sec:CentersOfHumpBump}

In RR~Lyrae stars, we observe that bumps occur on the descending branch of the light curve before the brightness minima, while the hump occurs on the ascending branch between the brightness minima and maxima. Using the centers of the humps and bumps for the stars in our sample we tested whether the positions of the shock features change with the pulsation properties. 

\begin{figure*}
\centering
\includegraphics[width=2\columnwidth]{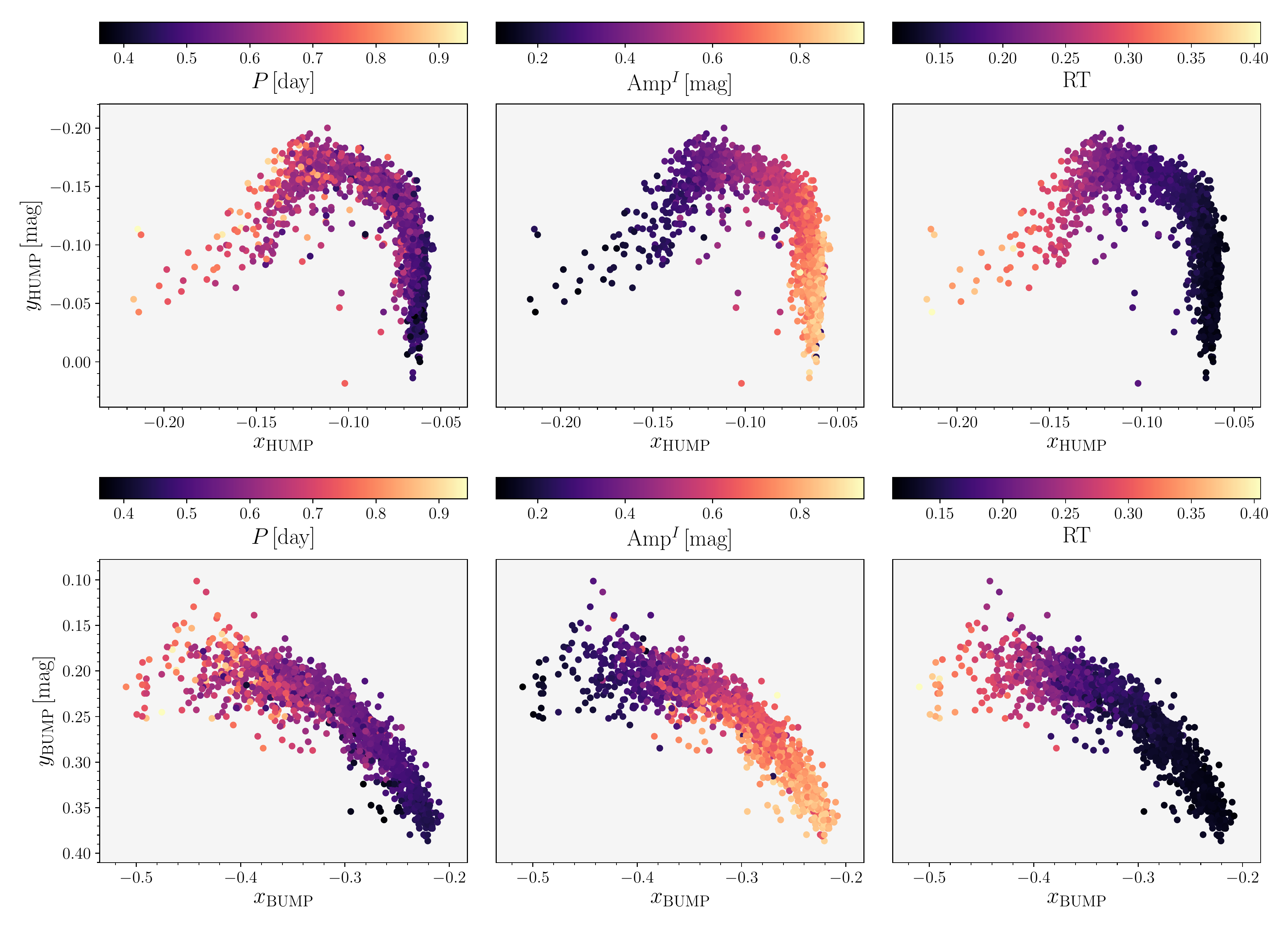} 
\caption{The relations between the coordinates of the centers for the main and early shock features in the phased light curves. The top panels display $x_{\rm HUMP}$ vs. $y_{\rm HUMP}$ and the bottom panels show the same dependence but for the bump, $x_{\rm BUMP}$ vs. $y_{\rm BUMP}$. The color-coding is based on the pulsation period (left-hand panels), total amplitude (middle panels), and RT (right-hand panels) of the individual stars.}
\label{fig:HumpsBumpsCenterAmp}
\end{figure*}

In Fig.~\ref{fig:HumpsBumpsCenterAmp} we displayed dependences between the coordinates of the humps (top panels) and bumps (bottom panels) in the normalized phased light curves. The color-coding of individual points is with respect to the pulsation period (left-hand panels), total amplitude of brightness changes (middle panels), and rise time (right-hand panels). The rise time (from here on referred to as RT) indicates the phase interval between the brightness minima and maxima on the phase curve. In these diagrams, we see that the centers of the humps occur on the ascending branch above the mean magnitude (negative values in $y_{\rm HUMP}$), contrary to bumps, where the centers appear on the descending branch below the mean magnitude (positive values in $y_{\rm BUMP}$). 

In the top panels, we see a correlation between the humps' center and the pulsation properties. The RRab variables follow the period-amplitude relation \citep[albeit with scatter caused, e.g. by the Oosterhoff dichotomy,][]{Oosterhoff1939}, thus the correlations with pulsation period and amplitude are most likely not independent. The location of the hump changes with pulsation period and amplitude: as we move from minimum periods and amplitudes toward the average ones, the hump moves toward the brightness maximum. An even stronger correlation with the pulsation properties is noticeable in the location of the bump. In this case, the bump moves toward the brightness minimum with increasing amplitude and decreasing pulsation period.

In the case of the horizontal coordinate $x_{\rm HUMP}$ and $x_{\rm BUMP}$ we used the RT (phase interval) for comparison. In the case of the $x_{\rm HUMP}$ we see a strong anti-correlation with the RT, with a Pearson correlation coefficient $r \approx -0.97$. The ratio of the two parameters clusters around 0.5, thus the hump occurs almost midway between the phases of brightness minimum and maximum. As in the case of the hump, we found a strong correlation between RT and $x_{\rm BUMP}$ for the early shock, with a Pearson correlation coefficient of $r \approx -0.89$ (for reference see right-hand panels of Fig.~\ref{fig:HumpsBumpsCenterAmp}). Thus, from Fig.~\ref{fig:HumpsBumpsCenterAmp} and from the aforementioned correlation coefficient we see that the pulsation properties are mainly correlated with the horizontal coordinate $x_{\rm HUMP}$ and $x_{\rm BUMP}$.

For the studied RR~Lyrae stars, we found that the location of the early and main shock on the phased light curves is connected with the pulsation properties of a given star. Although we note that pulsation period, amplitude, and RT are closely correlated; $r = -0.68$ for the pulsation period and amplitude, $r = -0.90$ for the amplitude and RT, and $r = 0.74$ for the pulsation period and RT. Therefore, the connection between the location of the humps and bumps with the pulsation parameters can stem from the correlation between individual pulsation parameters. 

\subsection{Relationships between shocks and light curve parameters} \label{sec:ShocksLightParam}

In this section, we will explore the relationships between the humps and bumps and the light curve properties of RR~Lyrae stars.

\begin{figure*}
\centering
\includegraphics[width=523.5307pt]{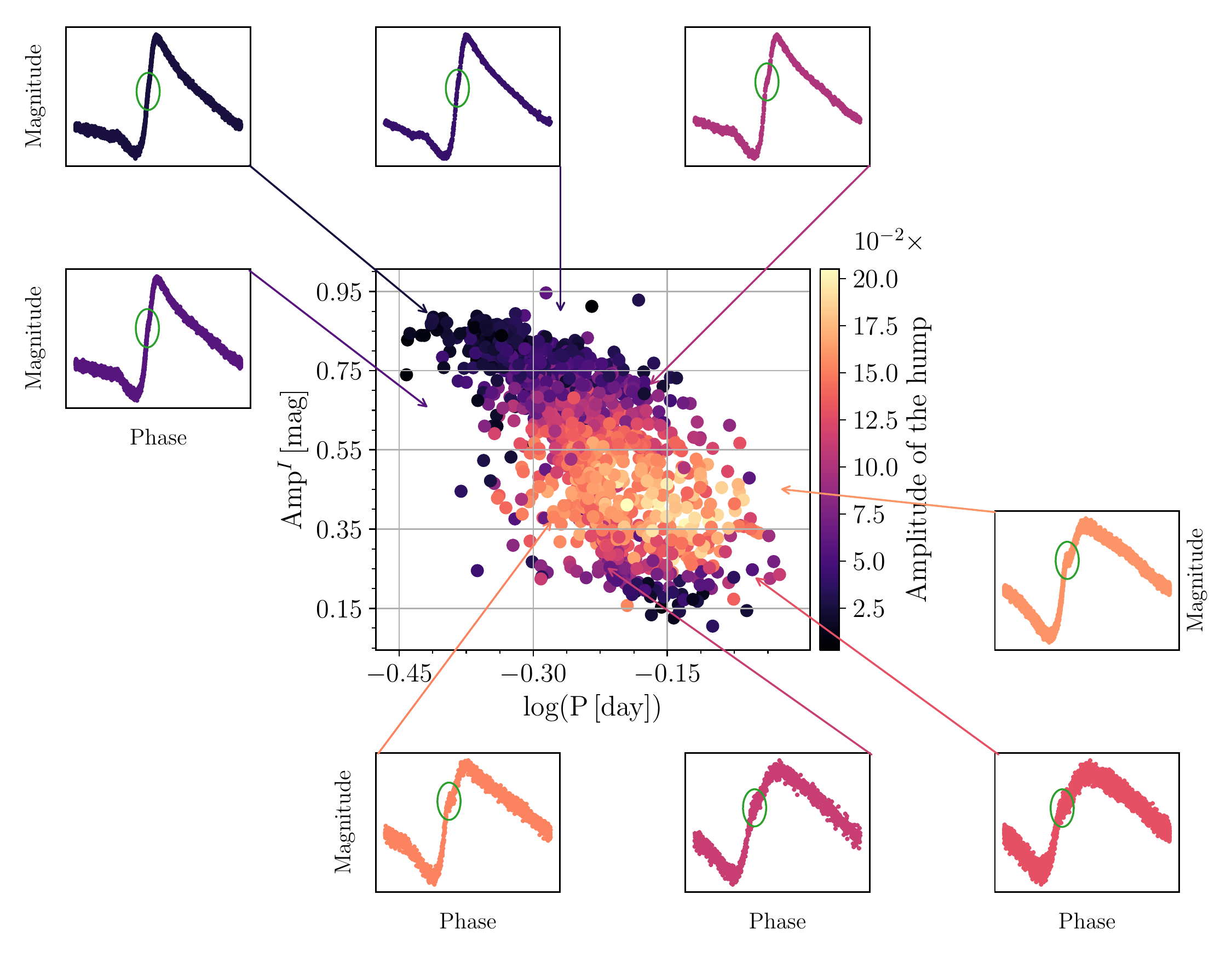}
\caption{The \textit{P-A} diagram for our studied stars with a color-coding based on the amplitude of the hump for the individual objects. The insets depict examples of phased light curves for a given region (grey grid) in the \textit{P-A} diagram indicated by the arrows. The green ellipses approximately denote the location of the hump.}
\label{fig:HumpsPAmap}
\end{figure*}

\begin{figure*}
\centering
\includegraphics[width=523.5307pt]{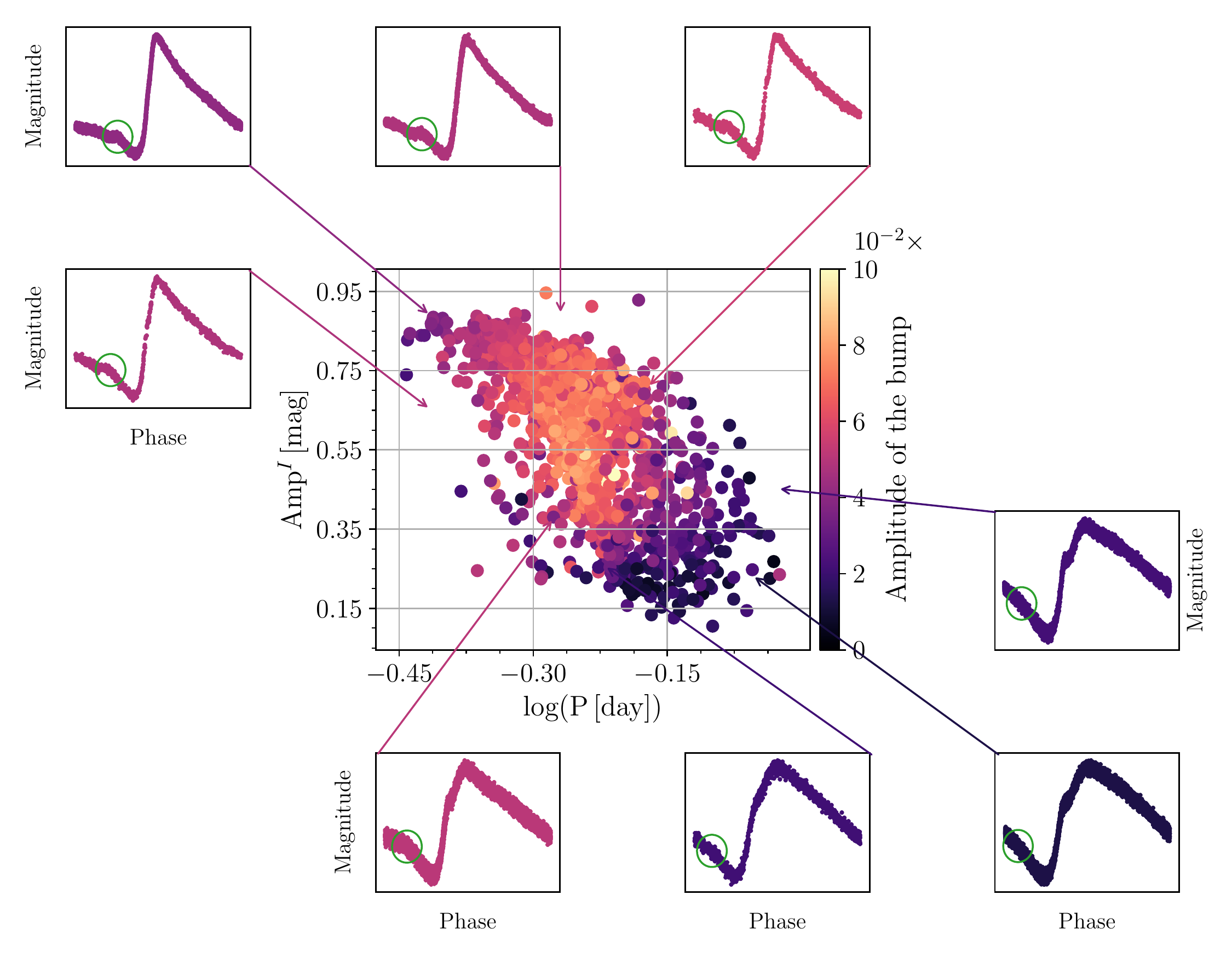}
\caption{Similar as Fig.~\ref{fig:HumpsPAmap} but the color-coding of the individual points is based on the amplitude of the bump, and the green ellipses approximately denote the location of the bump.}
\label{fig:BumpsPAmap}
\end{figure*}

One of the most important diagrams in studies focused on the RR~Lyrae stars is the period-amplitude (hereafter \textit{P-A}) diagram, which relies on reddening-free quantities and serves for, e.g., the separation of RR~Lyrae subtypes, identification of Oosterhoff populations, and rough estimates on the metallicity of their host stellar system. Fig.~\ref{fig:HumpsPAmap} shows the \textit{P-A} diagram with color-coding for the amplitude of the hump. Stars with the most pronounced main shock (hump) are located in the lower part of the \textit{P-A} diagram. We see a significantly less pronounced hump in the high-amplitude regime. The hump almost disappears at the very low-amplitude end of the \textit{P-A} diagram. The light curves of these stars do not appear to be affected at all by the main and early shocks (see Fig.~\ref{fig:BumpsPAmap}). The main shock has a negligible effect on the light curves of high-amplitude, short-period \citep[from here on referred to as HASP;][]{Fiorentino2015} RR~Lyrae stars. These regions in the diagram are marked by grey grid lines, and the arrows show to which region each of the insets around the \textit{P-A} diagram belong. The examples of phased light curves were chosen based on the average size of the amplitude of the hump in a given grid region of the \textit{P-A} diagram.

The situation for the bump in Fig.~\ref{fig:BumpsPAmap} is different. The effect of the early shock is stronger for stars with an amplitude higher than 0.4\,mag in the $I$~band. The HASP RR~Lyrae stars, contrary to the case of the main shock, show a pronounced effect of the early shock on the light curve. We see a nearly linear decline of the size of the bump with the decreasing amplitude and increasing pulsation period. A similar linear decrease is also visible if we move toward the HASP stars but variables in this region of the \textit{P-A} diagram still show a considerable hump size. Bumps among the low-amplitude (around 0.2\,mag in $I$~band) RR~Lyrae stars have almost no sign of an early shock in their light curves, similarly to the humps in Fig.~\ref{fig:HumpsPAmap}. 

\begin{figure*}
\centering
\includegraphics[width=\columnwidth]{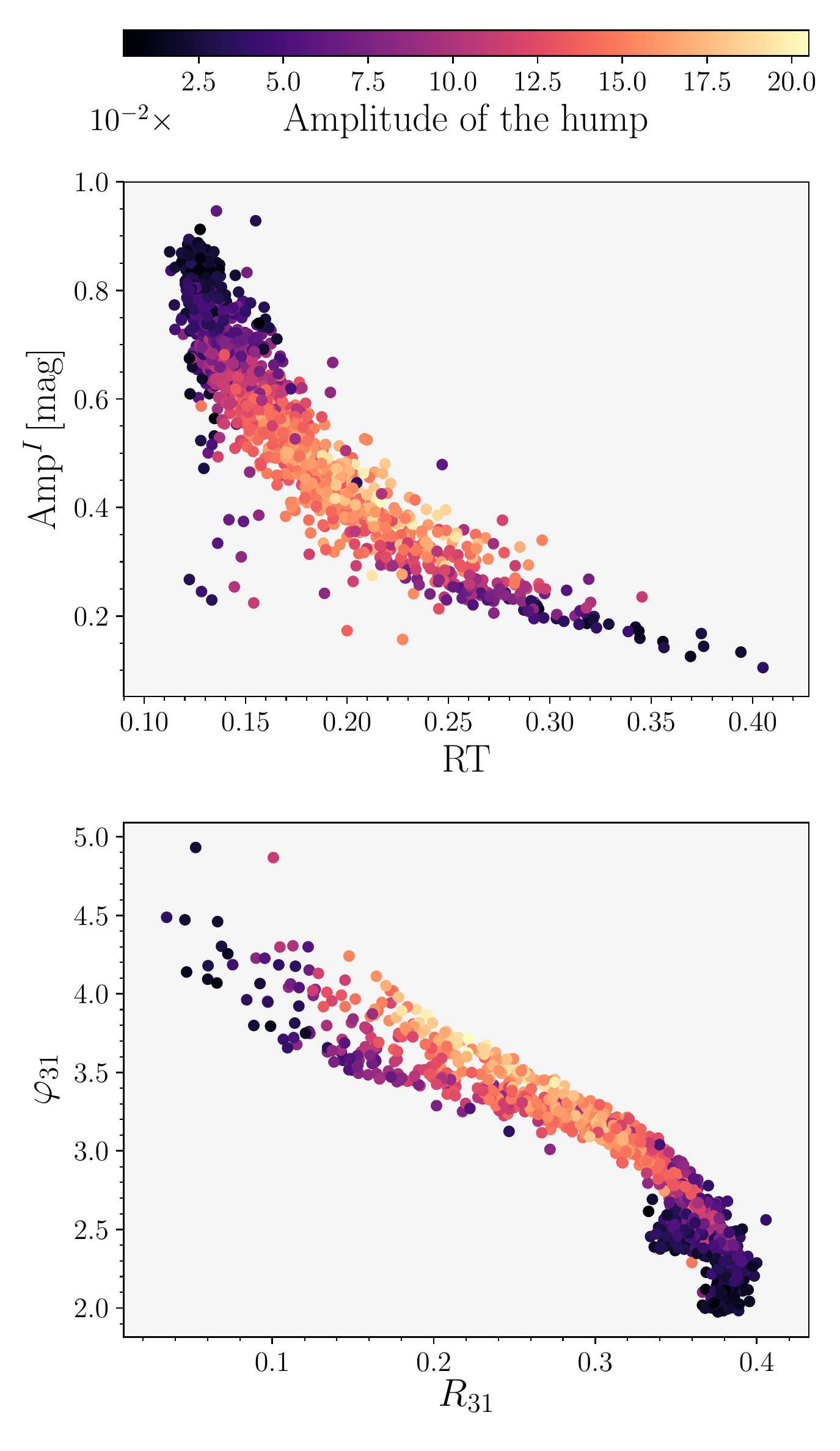}
\includegraphics[width=\columnwidth]{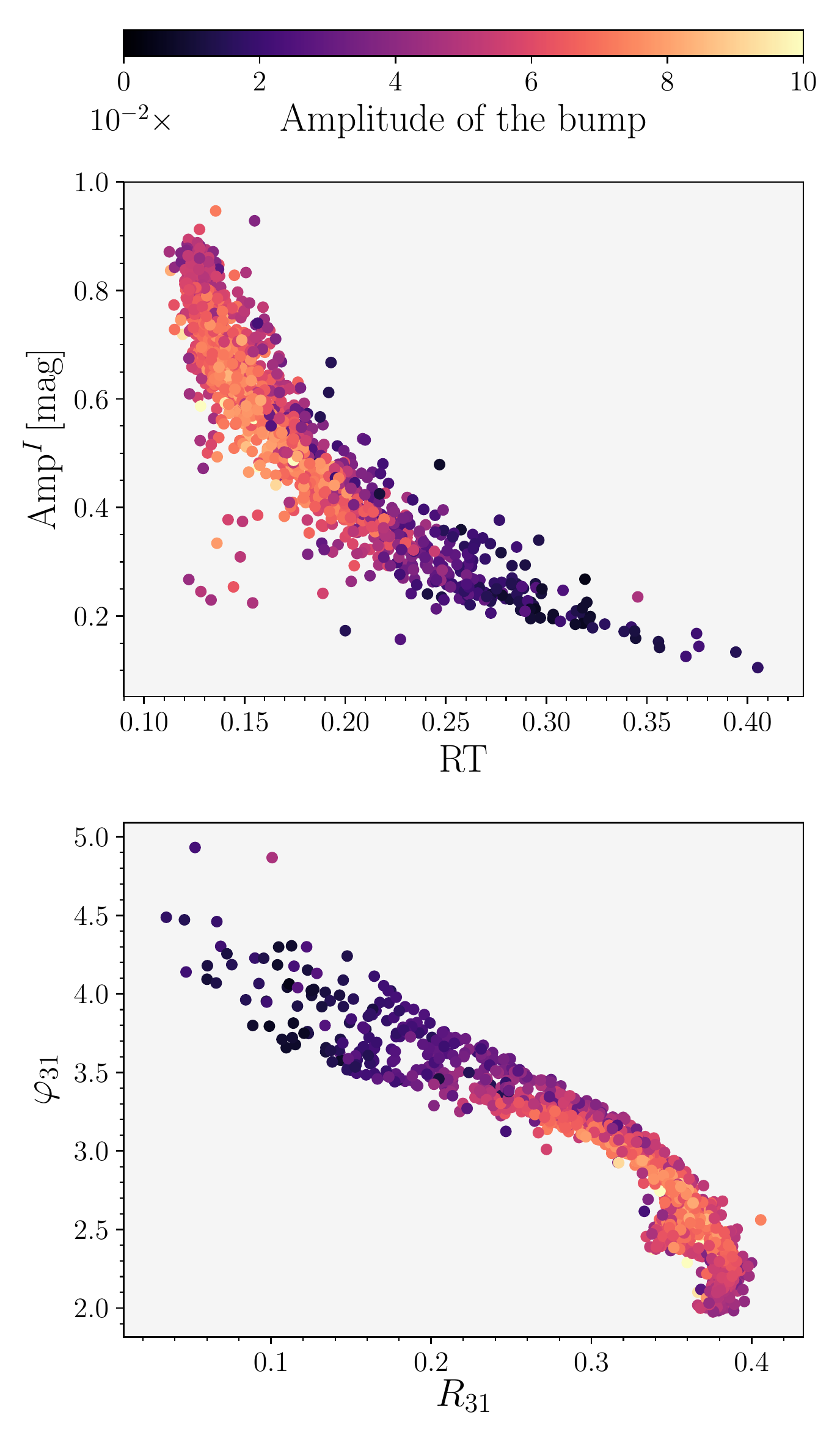}
\caption{The RT vs. pulsation amplitude (top panels) and $\varphi_{\rm 31}$ vs. $R_{\rm 31}$ (bottom panels), color-coded with respect to the size of the amplitude of the shock event.}
\label{fig:appParam}
\end{figure*}

In the top panels of Fig.~\ref{fig:appParam} we show the amplitude vs. RT dependence with the amplitude of the hump and bump color-coded. In the case of the hump, we see that variables with a pronounced hump (hump amplitude larger than 0.12) concentrate below the average value for the amplitude (0.6\,mag). As we move outward from the center the strength of the main shock decreases. Similarly to the case of the \textit{P-A} diagram we see that stars with a swift (small RT) or rather slow (high RT) change between minimum and maximum brightness tend to have smaller values for the amplitude of the hump than the stars with average RT. In the case of the bump (the right-hand top panel of Fig.~\ref{fig:appParam}) we see a different dependence. Stars with a pronounced early shock have a short RT and large amplitude (as shown in the case of the \textit{P-A} diagram). 

The bottom panels of Fig.~\ref{fig:appParam} show $\varphi_{\rm 31}$ vs. $R_{\rm 31}$ dependencies with a color-coding based on the strength of the hump (left-hand panels) and bump (right-hand panels). These diagrams can be used to identify modulated fundamental-mode RR~Lyrae stars \citep{Prudil2017Blazhko} or to identify signatures of the two Oosterhoff populations \citep{Prudil2019OO}. In the left-hand panel, we see that a large number of stars, with significant hump, is concentrated on the tail of the Oosterhoff\,II locus \citep[cf. top panel of fig.~4 in][]{Prudil2019OO}. In the right-hand panels the situation is different: stars with pronounced early shock are concentrated in the lower values of $\varphi_{\rm 31}$. The lower right panel shows that the increasing size of the bump is followed by an increase in $R_{\rm 31}$ and a decrease in $\varphi_{\rm 31}$. A more detailed analysis of the physical properties (e.g. metallicity, absolute magnitude, and effective temperatures) and association with Oosterhoff populations can be found in Sec.~\ref{sec:ShocksBigPicture}. It is important to point out that pulsation period, amplitude, RT, and other light curve parameters are strongly correlated, therefore it is expected to find correlations with other parameters as well. 

In Table~\ref{tab:PearsonCorrelation} we list the Pearson correlation coefficients, $r$, between the hump/bump sizes and pulsation properties of the studied stars. We highlight the strong correlations and anti-correlations between pulsation parameters and shock events. We note that the size of a hump is linked with some of the pulsation properties, namely the phase differences $\varphi_{\rm 21}$, $\varphi_{\rm 31}$, and $I$ band amplitude. Furthermore, for the amplitude ratios $R_{\rm 21}$ and $R_{\rm 31}$ we observe negligible anti-correlations with the main shock. For the early shock, we observe strong and moderate correlations between the prominence of the bump and most of the pulsation properties. We also notice that in almost all cases we observe an anti-correlation between the sizes of a hump and bump for individual pulsation properties. In cases where we observe a positive correlation with the hump and the pulsation feature, we observe a negative correlation for the bump. We searched for correlations among higher orders of the Fourier amplitude ratios (up to $R_{\rm 151}$) and found that the strongest correlations with the main shock are for $R_{\rm 61}$, $R_{\rm 131}$, and $R_{\rm 141}$, while the early shock is more correlated with lower orders -- $R_{\rm 21}$, $R_{\rm 31}$. To summarize, several of the pulsation properties of the studied RR~Lyrae stars seem to be closely connected with the shock effects on the phased light curves.

\begin{table}
\caption[]{Table for the Pearson coefficients $r$ between various pulsation properties and the effect of the main and early shock on the phase curves. The columns from left to right list the pulsation features, columns 2 and 3 list Pearson coefficients. Strong correlations $\left( \left | r  \right |> 0.7 \right)$ are highlighted in boldface fonts.} %
\label{tab:PearsonCorrelation}
\begin{tabular}{lcc}
\hline \hline       
& $r^{\rm HUMP}$ & $r^{\rm BUMP}$ \\
\hline
$R_{\rm 21}$ & $-0.182$ & $\textbf{0.709}$ \\
$R_{\rm 31}$ & $-0.390$ & $\textbf{0.717}$ \\
$\varphi_{\rm 21}$ & $0.667$ & $-0.560$ \\
$\varphi_{\rm 31}$ & $0.641$ & $-0.563$ \\
P\,[day] & $0.537$ & $-0.466$ \\
Amp$^{I}$\,[mag] & $-0.630$ & $0.449$ \\
Amp$^{K}$\,[mag] & $-0.388$ & $0.389$ \\
RT & $0.425$ & $-0.661$ \\
\hline
\end{tabular}
\end{table} 

We note that, in the available light curves, we do not see other shock events \citep[e.g., a \textit{jump}, or \textit{lump};][]{Chadid2014}, but in a handful of cases we notice a feature before the bump that can be ascribed to the post-maximum shock wave or rump \citep{Chadid2014}.

\begin{figure}
\centering
\includegraphics[width=\columnwidth]{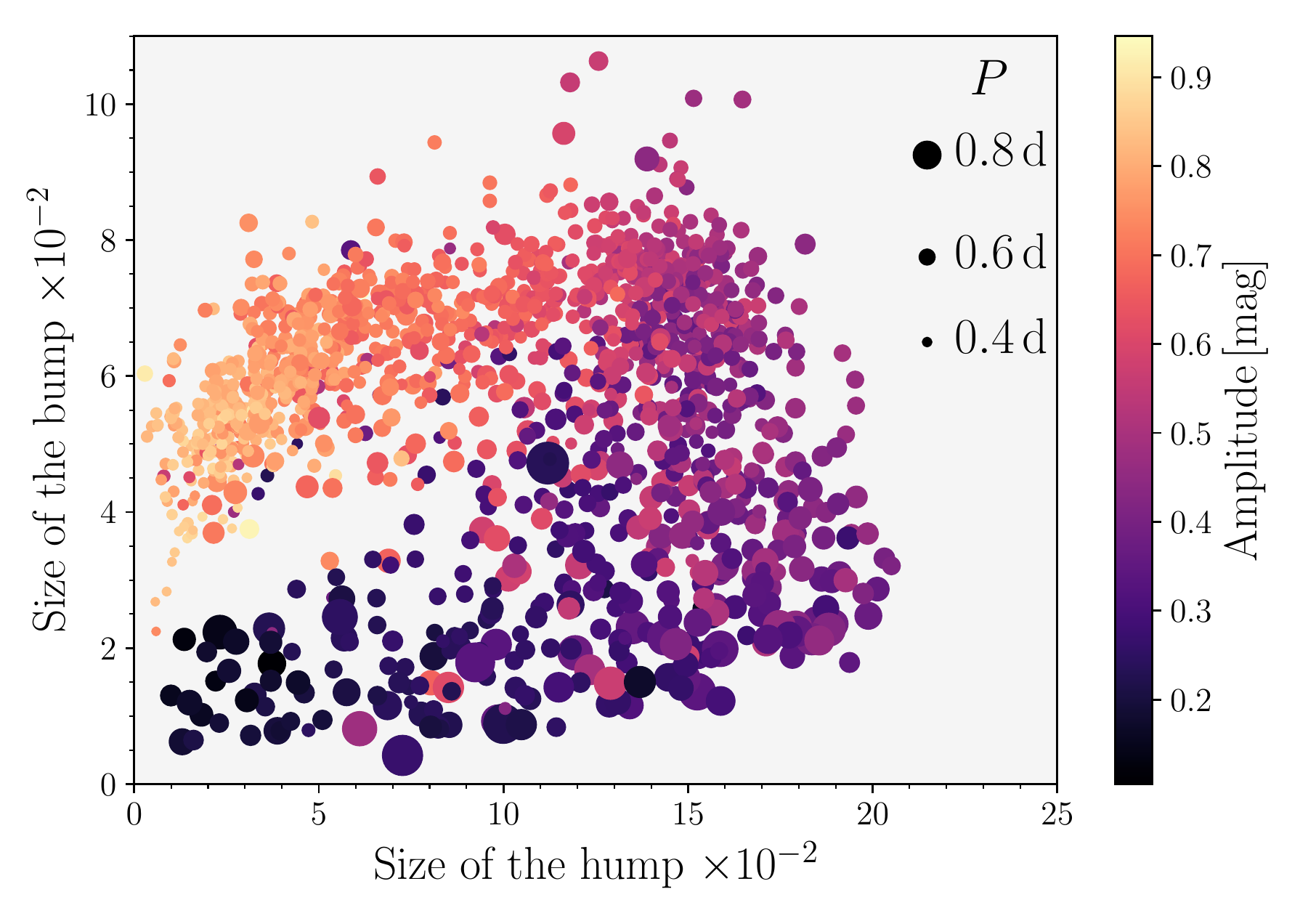} 
\caption{The size distribution of main and early shock. The color-coding and point size are based on the amplitude and pulsation period of the individual objects as shown in the legend.}
\label{fig:humpbumpfour}
\end{figure}

The link between the pulsation parameters and the sizes of the hump and bump is also visible in Fig.~\ref{fig:humpbumpfour}, where we compare the sizes of the humps and bumps in our RR~Lyrae sample. We see that stars with the highest amplitudes (brighter points) and short periods (smaller points) have an average size early shock and almost no main shock feature in their phased light curves. On the other hand, we see that the majority of the stars with the lowest amplitudes (below 0.2\,mag) have very little signs of both shocks. Unlike for short period RR~Lyrae stars, the pronounced main shock seems to be concentrated among the average and long-period RR~Lyrae variables.

\section{Shocks in the color-magnitude diagram} \label{sec:ShocksBigPicture}

This section is focused on the connection between the sizes of the main and early shock with general properties of the studied stars from \citet{Prudil2019OO} (metallicities, Oosterhoff associations, color indices and dereddened magnitudes).

For the purpose of this paper, we used the Oosterhoff identification provided by \citet{Prudil2019OO}; from our studied RR~Lyrae sample, 1101 objects are associated with the Oosterhoff group I (Oo\,I) and 384 belong to the Oosterhoff group II (Oo\,II). We compared whether the appearance of humps and bumps in the two Oo populations is different. We noticed that the size of the main shock is the same within the errors for both Oosterhoff groups. For the Oo\,II group the median value of the amplitude of the hump is equal to $(12 \pm 5) \cdot 10^{-2}$, and for the Oo\,I population its value is $(10 \pm 5) \cdot 10^{-2}$. The median value for the amplitude o the bump is again similar for both Oo types, $(6 \pm 1) \cdot 10^{-2}$ (Oo\,I stars) and $(5 \pm 2) \cdot 10^{-2}$ (Oo\,II stars), which is in agreement with the population effect of the Oosterhoff dichotomy as suggested by \cite{Fabrizio2019}.

\begin{figure*}
\centering
\includegraphics[width=\columnwidth]{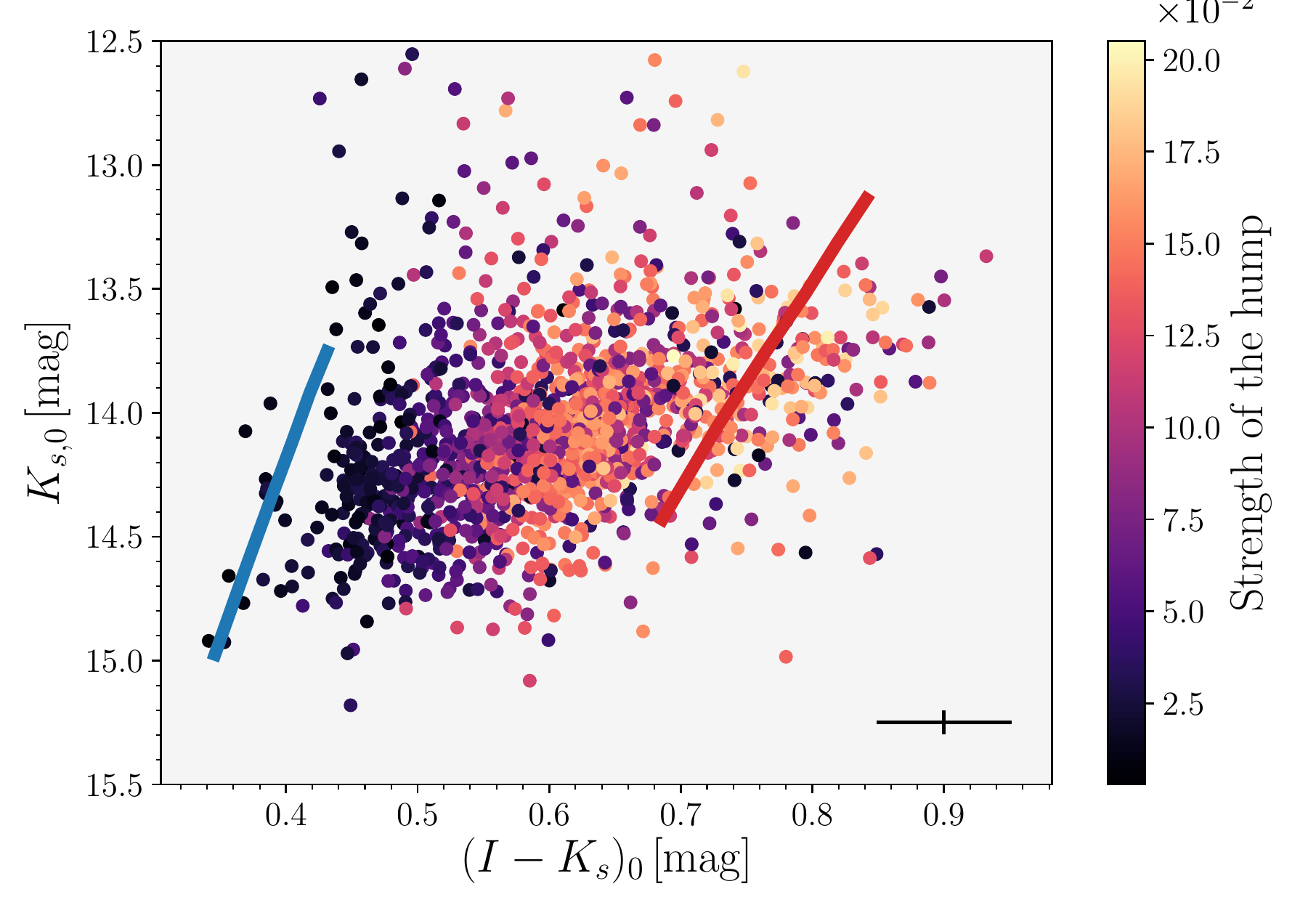}
\includegraphics[width=\columnwidth]{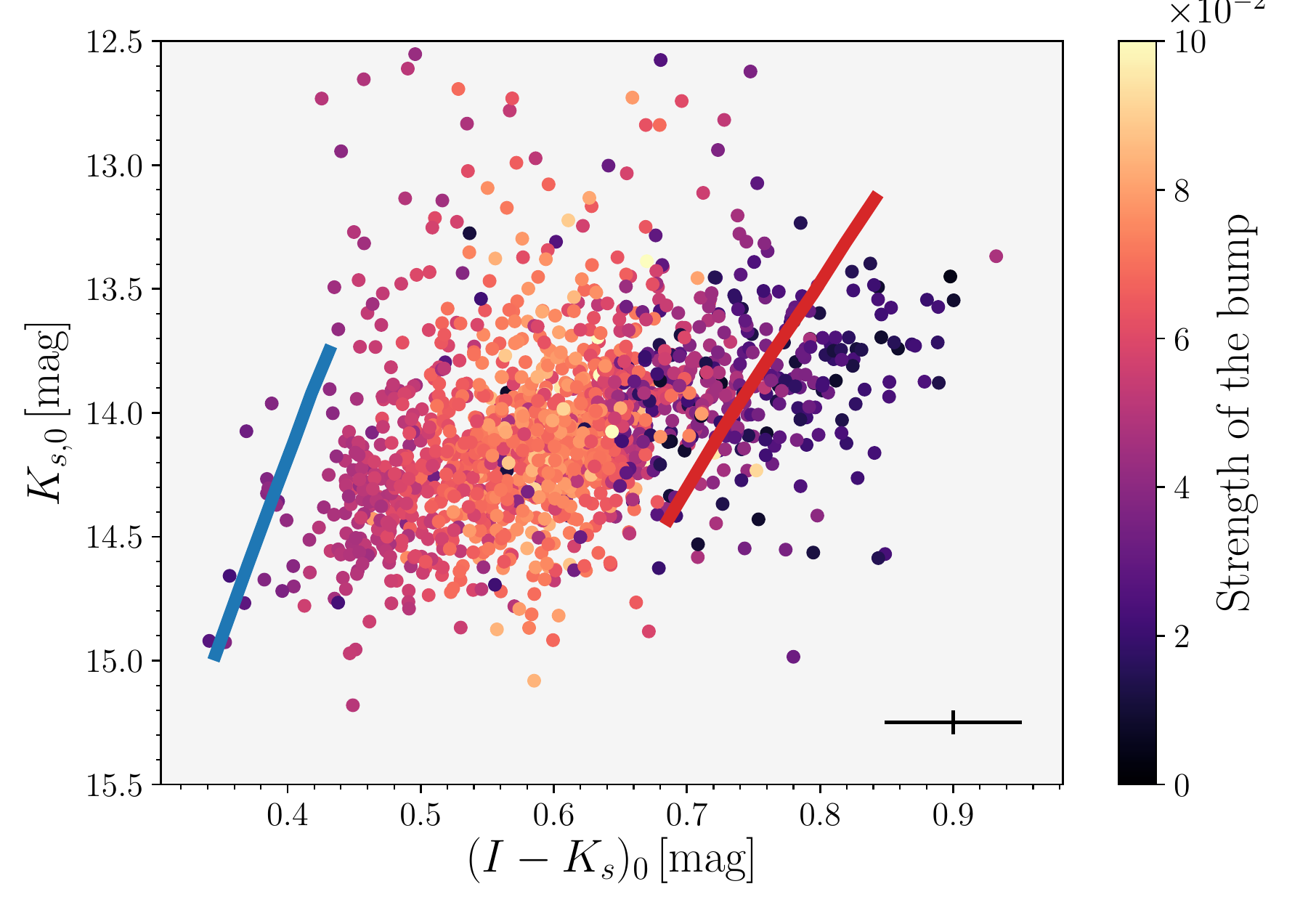}
\caption{The CMD for the studied stars with color-coding representing the effect of the hump (left-hand panel) and bump (right-hand panel). The edges of the fundamental-mode IS are marked with blue and red solid lines that were calculated using the linear pulsation models from \citet{Smolec2008}. The median of the errors on color and mean magnitude are depicted with black lines in the bottom right corner of each panel.}
\label{fig:HumpsBumpsVsCMD}
\end{figure*}

\begin{figure}
\centering
\includegraphics[width=\columnwidth]{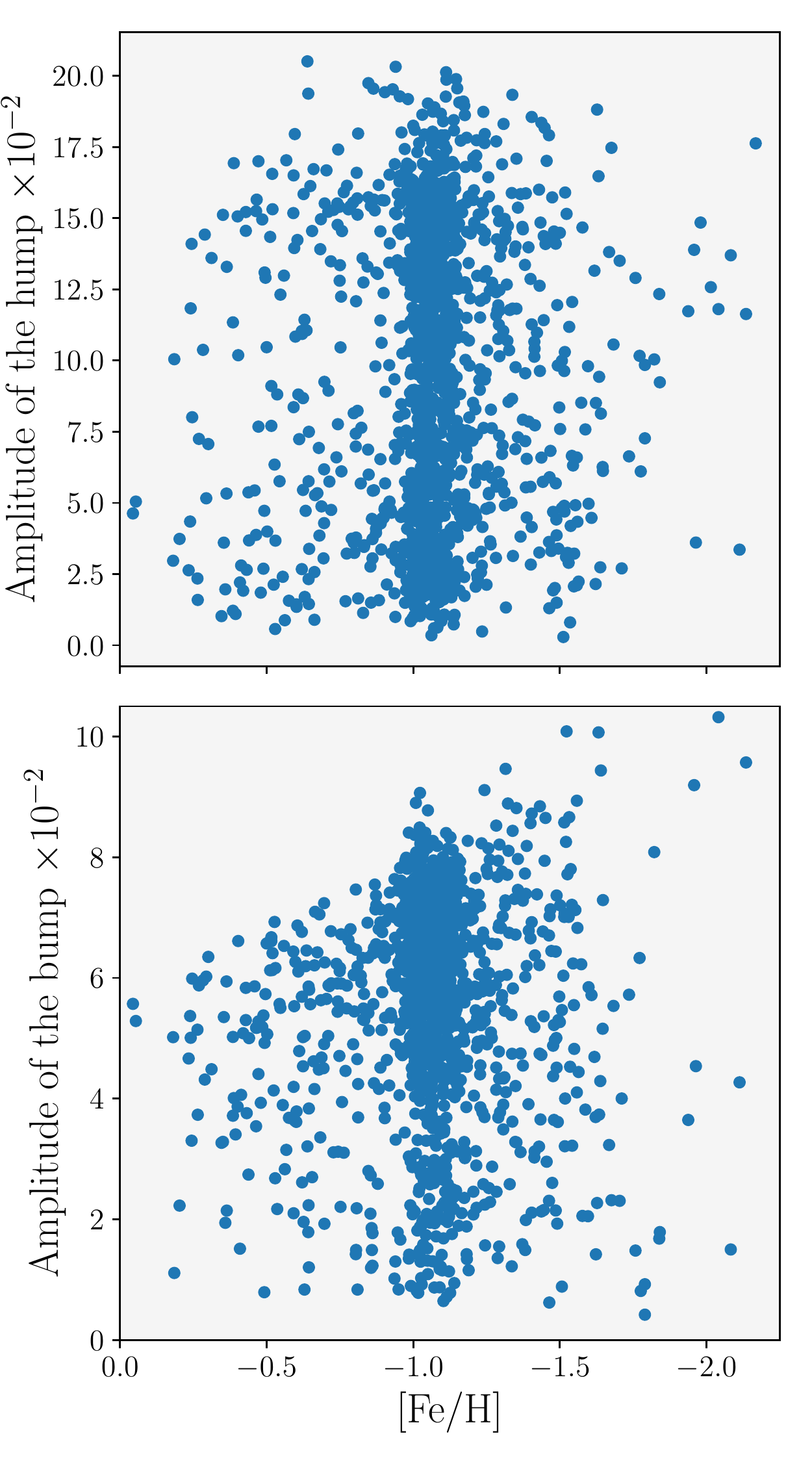}
\caption{The photometric metallicity [Fe/H] is shown as a function of the hump (top panel) and bump (bottom panel) strength for all studied RR~Lyrae stars.}
\label{fig:HumpsBumpsVsFEH}
\end{figure}

The color-magnitude diagram (from here on CMD), especially the fundamental mode instability strip (IS) of the studied stars, Fig.~\ref{fig:HumpsBumpsVsCMD}, shows very interesting features. Stars with a strong hump are concentrated on the cooler side of the IS, and as we move toward the blue edge of the IS, the hump becomes less pronounced and almost vanishes around $\left(I - K\right)_{0} \approx$ 0.55\,mag. Stars that we find close to the blue edge of the fundamental-mode IS are stars with high amplitudes and short pulsation periods (see {\it P-A} diagram in Fig.~\ref{fig:BumpsPAmap}). We note that we find stars with very little sign of the hump at the red edge of the IS, but they are scarce. For the bump the situation is almost the opposite, we see stars with a pronounced bump in the blue and central part of the IS. As we move toward the fundamental red edge, the effect of the bump on the phased light curve decreases, and almost vanishes once we cross $\left(I - K\right)_{0} \approx$ 0.7\,mag. This correlation between the position in the CMD and the shocks is most likely not connected with the metallicity, since we detect no correlation between [Fe/H] and the shocks ($r=-0.05$ for humps and $r=-0.08$ for bumps, respectively, see Fig.~\ref{fig:HumpsBumpsVsFEH}), but appears to be connected with the pulsation periods (for correlation coeficients see Tab.~\ref{tab:PearsonCorrelation}). In the CMDs we noticed a smooth gradient of the bump strength, unlike in the case of the hump. This is most likely due to the higher intrinsic scatter of the hump properties, and stars that have a pronounced hump (e.g., 0.14 and higher) are much less numerous than stars that have pronounced bump (e.g., 0.05 and higher) by at least a factor of 2. We simply do not have that many stars with amplitudes around 0.45\,mag, and periods of 0.7\,days (where the hump is large), but we have plenty of stars with amplitudes around 0.55\,mag and periods around 0.55\,days (where the bump is large).

In addition, the strong correlation with colors and amplitudes suggests that we are looking at the combination of these effects in the CMD. The horizontal axis of the CMDs in Fig.~\ref{fig:HumpsBumpsVsCMD} is based on the period metallicity luminosity relations; \citep{Catelan2004,Muraveva2018}; furthermore, the same relations were used to estimate the extinction in the direction of the studied stars. Thus these assumptions enter into our analysis and the general properties from the {\it P-A} diagram transfer into the CMD distribution of the RR~Lyrae stars. 

We note that the lines in Fig.~\ref{fig:HumpsBumpsVsCMD} (blue and red) depicting the boundaries of the IS for the fundamental mode pulsators were calculated using linear pulsation models from \citet{Smolec2008} with a mixture of heavy elements based on \citet{Asplund2009} and OPAL opacities \citep{Iglesias1996}. We used the same grid of physical parameters as in \citet{Prudil2019OO}. 

\subsection{The period change rate} \label{sec:PeriodChange}

In the left panel of Fig.~\ref{fig:HumpsBumpsVsCMD} we see that some stars without observable hump are present close to the red edge of the IS, therefore in a region where we also find stars with a rather pronounced hump (0.5 -- 0.7\,mag range on the horizontal axes). To gain further insight into this phenomenon, we have investigated whether the prominence of the hump and bump features may be connected with the direction of the evolution of the star in the CMD. At least in principle, one may assess the evolutionary direction of a pulsating star by means of the secular rate of change of its period, a phenomenon that is caused by the slow change in the star's mean density as it crosses the IS. In general, an RR Lyrae star that crosses the IS on its way to the asymptotic giant branch (i.e., redwards) is expected to show a positive period change rate and vice-versa for those evolving away from the zero-age horizontal branch (ZAHB) in blueward direction.

In order to explore the possible connection between shocks in RR~Lyrae atmospheres and their evolutionary status, we used the data from previous releases of the OGLE survey of the Galactic bulge \citep{Udalski1994I,Udalski1995III,Udalski1995II,Udalski1996IV,Udalski1997V,Soszynski2011OGLEIII,Soszynski2014OGLEIV}. The OGLE photometry spans up to 25 years for some objects, which provides a more or less sufficient baseline to study changes in the pulsation period. Some of our sample stars were not observed during the previous phases of the OGLE survey, therefore they have a substantially smaller baseline ($\approx$ 7 years), and we omitted them from the subsequent analysis (46 objects). 

\begin{figure}
\centering
\includegraphics[width=\columnwidth]{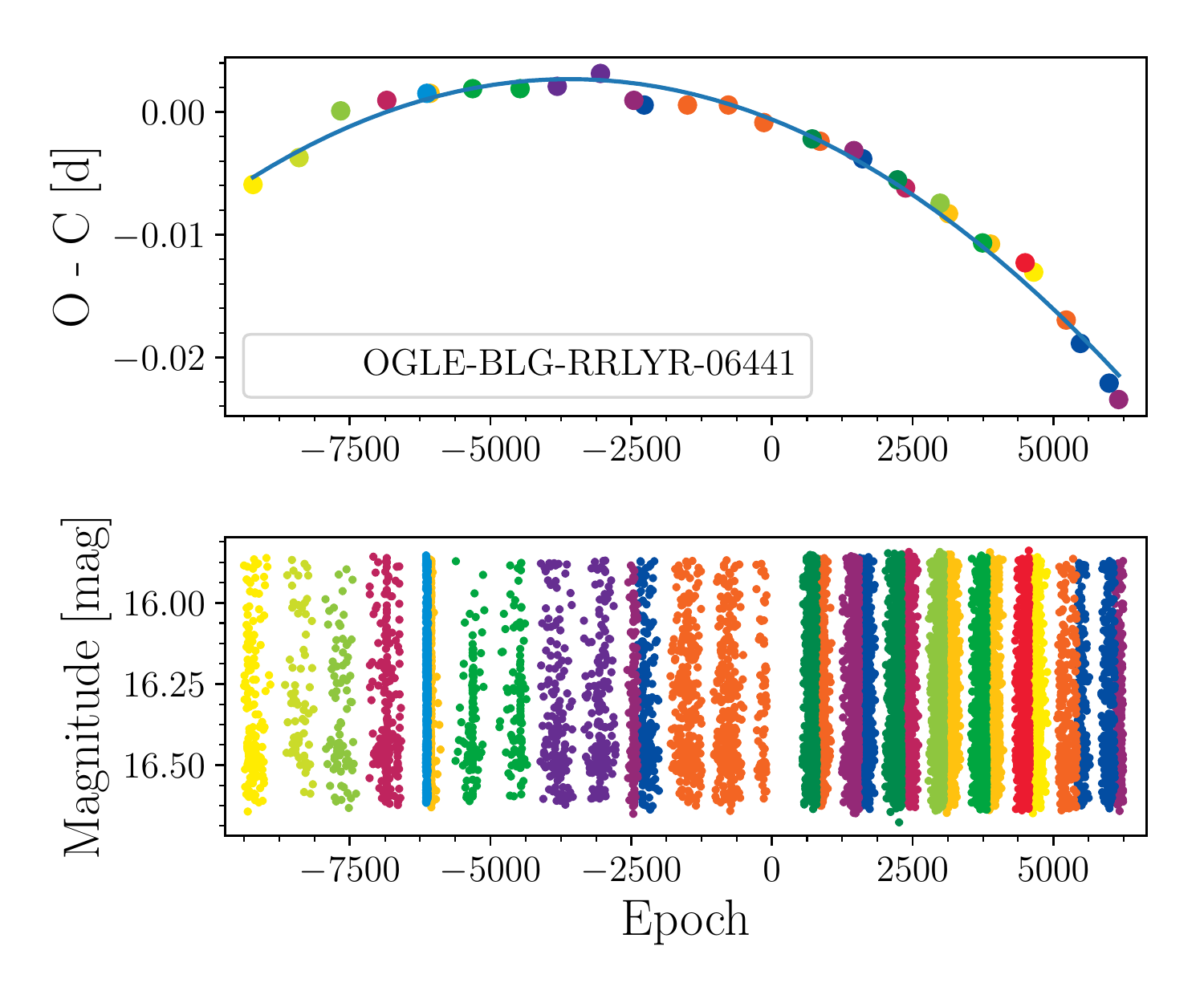}
\caption{An example of an $O - C$ diagram for one of the stars from our sample (top panel) with all available photometry in the bottom panel. The photometry in the bottom panel is color-coded based on the data bins for the construction of the $O - C$ diagram.}
\label{fig:6441-OC}
\end{figure}

For the purpose of constructing the $O - C$ diagrams (observed minus calculated) for the studied RR~Lyrae stars, we utilized the whole light curves using the method proposed by \citet{Hertzsprung1919}, where we create a phased light curve template that we used to estimate the phase shift between individual observational seasons. We binned the available photometry based on the observational seasons (color-coding in the bottom panel of Fig.~\ref{fig:6441-OC}). For each subsample we phased the photometric data using the pulsation ephemerides from the OGLE-III data release. Each phased subsample was then fitted with a light curve template. The shift in phase of the brightness maximum was estimated and multiplied by the pulsation period in order to get the $O - C$ value (depicted in the top panel of Fig.~\ref{fig:6441-OC} with respect to the time span). 

The $O - C$ diagrams of RR~Lyrae stars can show various trends. The most common ones are a linear trend across the epochs (inaccurate ephemerides) or a parabolic trend that suggests a change in the pulsation period. To estimate possible changes in the pulsation periods, the data in the $O - C$ diagram are fitted with the following quadratic equation:
\begin{equation}
O - C = a_{0} + a_{1}\cdot E + a_{2}\cdot E^{2},
\end{equation}
where $a_{n}$ are coefficients of the polynomial function of degree $n$, and $E$ is the number of epochs that elapsed from $M_{0}$ (in our case $M_{0}$ corresponds to an initial time of maximum brightness from the OGLE-III survey). The period change rate $\beta$ is calculated via:
\begin{equation}
\beta = \frac{2 \cdot a_{2}}{P}.
\end{equation}
Not all of the stars from our sample undergo a period change; some have simply incorrectly determined ephemerides. In order to separate stars with changing pulsation periods from those with constant periods, we employed the same criterion as used by \citet{Jurcsik2001}:
\begin{equation}
\left | a_{2} \right | / \sigma_{a_{2}} > 2.
\end{equation}
Stars that fulfill this condition were considered as pulsators undergoing an increase/decrease in their pulsation periods. For 956 stars from our sample, we detected changes in the pulsation periods. It is important to add that the use of the period-change rate as an evolutionary clock for RR~Lyrae stars is hampered by the period change noise (for more details see Subsection \ref{subsec:NoteOnOoster}). 

\begin{figure}
\centering
\includegraphics[width=\columnwidth]{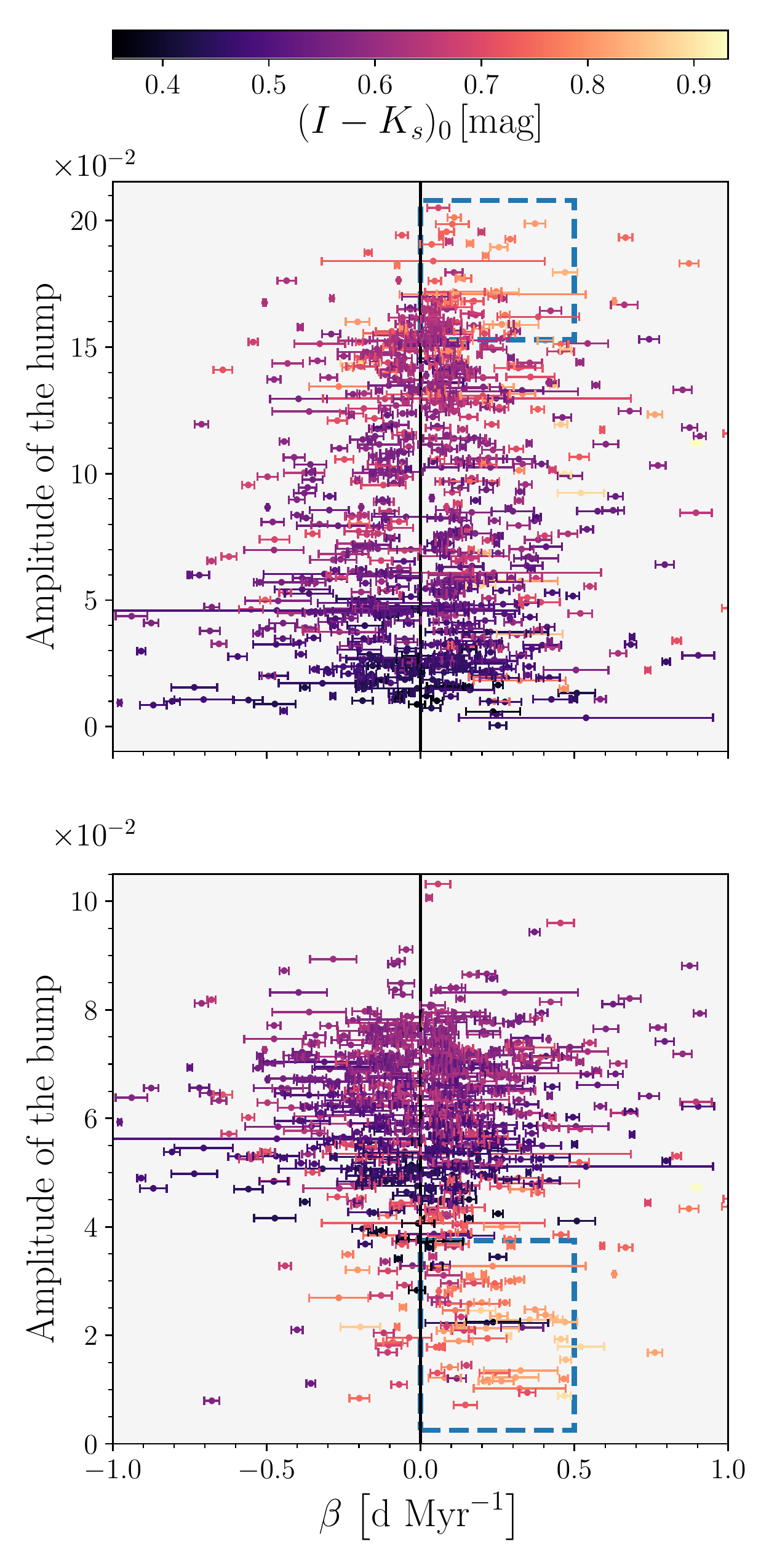}
\caption{The period change rate $\beta$ is shown as a function of the size of the amplitude of the hump (top panel) and bump (bottom panel). The color coding is based on the color $\left(I - K_{s}\right)_{0}$. The two blue dashed rectangles in both panels mark the positions of two groups that lie close to the red edge of the IS.}
\label{fig:period-change}
\end{figure}

In Fig.~\ref{fig:period-change} we show the dependences of the period change rate on the strength of humps (top panel) and bumps (bottom panel) and their color, $\left(I - K_{s}\right)_{0}$. In both panels we see that the majority of the stars cluster inside the $\pm$0.5\,d\,Myr$^{-1}$ range and approximately 10\,\% of the stars have a higher period change rate $\beta$, of which 66\,\% show an increase in pulsation periods \citep[similar to the previous studies of period changes in RR~Lyrae stars; e.g.,][]{LeBorgne2007,Kunder2011}. We do not observe any differences in $\beta$ in stars close to the red edge of the IS with different shock sizes. What we see in general, especially in the bottom panel, is that stars located near the red edge of the IS have on average a larger $\beta$ in comparison with stars in the middle or close to the blue edge of the IS. Stars with large positive $\beta$ in the red part of the IS are probably leaving the IS (marked with a dashed blue rectangle in Fig.~\ref{fig:period-change}), while stars with large negative $\beta$ are probably just arriving at the IS and ZAHB \citep{Silva-Aguirre2008} or are undergoing sudden structural instabilities before leaving the IS \citep[e.g.,][]{Sweigart1979,Koopmann1994}. We do not observe any relation between the period change rate and the strength of the hump or bump with respect to their position in the IS.

In Appendix~\ref{sec:AppTable} we included the first lines of a table for our dataset. The table contains the photometric properties of the studied stars, their metallicities determined on the basis of the photometric data, information about the measured shock sizes, the stars' association with the Oosterhoff groups, and the period change rate. The table in full is included in the supplementary material.

\subsubsection{Note on the period change rate with respect to the Oosterhoff groups} \label{subsec:NoteOnOoster}

One of the possible explanations of the Oosterhoff dichotomy is the hysteresis mechanism, which suggests that Oo\,I stars evolve from the red edge toward the blue edge of the IS while Oo\,II stars evolve in the opposite direction \citep{vanAlbada1973}. This hypothesis has been tested by, e.g., \citet{Kunder2011}, who calculated the period change rate for RR~Lyrae stars in IC~4499 and found no correlation between $\beta$ and the Oosterhoff population. In general, we observe a positive period change rate among RR~Lyrae stars in various globular clusters. Therefore, we used our large sample to test this hypothesis based on the determined period change rates and association with the Oosterhoff groups for individual pulsators. We found that 59\,\% of the stars from our sample exhibit a positive period change, which is in agreement with the observations of MW globular clusters, see fig.~15 in \citet{Catelan2009} or fig.~6.16 in \cite{Catelan2015book}. Both Oosterhoff groups exhibit on average a positive period change rate, 0.02\,d\,Myr$^{-1}$ and 0.16\,d\,Myr$^{-1}$ for Oosterhoff\,I and Oosterhoff\,II, respectively. We find that 68\,\% of the Oo\,II and 55\,\% of the Oo\,I RR~Lyrae stars have a positive period change rate. In addition, the vast majority of stars with significant period change rates (positive and negative) are associated with the Oo\,II group and are located close to the red edge of the IS. 

We note that the findings above can be affected by the presence of peculiar period changes (period change noise) with respect to the theoretical predictions. For RR Lyrae stars with the largest observed period change rates, deviations with respect to theoretical predictions can reach up to an order of magnitude or more \citep[see also][]{Jurcsik2012}. Furthermore, some RR~Lyrae undergo large, sudden changes in their pulsation periods, the latter sometimes increasing but others decreasing \citep[see, e.g.,][]{Catelan2015book}, and there is at least one known case of a double-mode RR Lyrae star (V53 in M15) whose fundamental and first overtone components have period change rates of different signs \citep{Paparo1998}. There have been several attempts to explain the presence of period change noise, invoking, e.g., large mass loss episodes \citep{Laskarides1974}, discrete mixing events \citep{Sweigart1979}, the difference in the helium composition below the convective zone \citep{Cox1998}. Considering the timespan of our sample, some of the studied stars may be affected by the aforementioned abrupt changes in pulsation periods and affect the overall conclusions.

\section{Summary and conclusions} \label{sec:Conclus}

In this paper, we provide the most extensive study of the effects of the main and early shock on the light curves of RR~Lyrae stars. We used photometric data from the fourth data release of the OGLE Galactic bulge survey for fundamental mode RR~Lyrae variables. We selected only non-modulated variables with abundant and good quality photometry studied in \citet{Prudil2019OO}. In the end, we analyzed 1485 RR~Lyrae stars and estimated the location and size of the humps and bumps in the phased light curves. 

We found that the positions of the humps and bumps in the phased light curves are related to the pulsation properties (pulsation period, amplitude, RT) of the studied stars. Main and early shocks are most pronounced in different regions of the \textit{P-A} diagram. Stars with a pronounced hump are centered around average and below-average amplitudes at longer pulsation periods. As we move toward variables with rather low/high amplitudes, the main shock weakens and eventually disappears from the phased light curves, especially for HASP variables \citep{Fiorentino2015}. The pulsators with a pronounced early shock are, on the other hand, located in the high amplitude regime. With decreasing amplitude, the effect of the bump declines. Overall, the bump size seems to strongly correlate with the amplitude ratios of the studied stars.

The CMD of the studied stars revealed a dichotomy between stars with and without strong shock features. This dichotomy is probably caused by the correlation between pulsation periods and amplitudes with colors and dereddened magnitudes, which are based on period-metallicity-luminosity relations. Further examination using the photometry from the TESS and {\it Kepler} space telescopes in combination with {\it Gaia} parallaxes could shed some light on the structure of the IS based on the humps and bumps. 

Several of our studied stars stand out in the CMD with their weak shock feature while being located in a region with a rather pronounced hump or bump. In order to investigate their evolutionary status, we calculated the period change rate $\beta$ from constructed $O - C$ diagrams. We found that the majority of the stars clump around period change rates of $\pm$0.5\,d\,Myr$^{-1}$ and only 10\,\% of the stars exceed this period change rate \citep[which is in agreement with previous studies of the period-change rate, e.g.,][]{LeBorgne2007,Kunder2011,Jurcsik2012}. We do not observe a difference in period change rate among stars with different strengths of the main and early shock. In general, the majority of the analyzed stars show positive $\beta$, which is especially evident close to the red edge of the IS, where stars that are most likely leaving the IS are expected. 

This is also reflected in the period change rate among the two Oosterhoff groups in our sample. In the Oosterhoff population II, 68\,\% of the stars have a positive period change, while in Oosterhoff group I, 55\,\% exhibit a period increase. This suggests that stars of Oosterhoff group I are located close to the ZAHB, while Oosterhoff population II variables already left the ZAHB and are moving toward the red side of the instability strip \citep[a similar effect was also observed in M3 by][]{Jurcsik2012}.

A dedicated spectroscopic survey in parallel with precise photometry optimized for shocks and their effects over a large spectral range should provide deeper insight into this phenomenon.

\begin{acknowledgements}
Z.P. acknowledges the support of the Hector Fellow Academy. R.S. was supported by the National Science Center, Poland, grant agreement DEC-2015/17/B/ST9/03421. I.D. and E.K.G were supported by Sonderforschungsbereich SFB 881 ''The Milky Way System'' (subprojects A02, A03, A11) of the German Research Foundation (DFG). Support for M.C. is provided by Fondecyt through grant \#1171273; the Ministry for the Economy, Development, and Tourism's Millennium Science Initiative through grant IC\,120009, awarded to the Millennium Institute of Astrophysics (MAS), and by Proyecto Basal AFB-170002.
\end{acknowledgements}

\bibliographystyle{aa}
\bibliography{biby} 

\begin{thebibliography}{71}
\expandafter\ifx\csname natexlab\endcsname\relax\def\natexlab#1{#1}\fi

\bibitem[{{Asplund} {et~al.}(2009){Asplund}, {Grevesse}, {Sauval}, \&
  {Scott}}]{Asplund2009}
{Asplund}, M., {Grevesse}, N., {Sauval}, A.~J., \& {Scott}, P. 2009, \araa, 47,
  481

\bibitem[{{Belokurov} {et~al.}(2018){Belokurov}, {Deason}, {Koposov},
  {Catelan}, {Erkal}, {Drake}, \& {Evans}}]{Belokurov2018}
{Belokurov}, V., {Deason}, A.~J., {Koposov}, S.~E., {et~al.} 2018, \mnras, 477,
  1472

\bibitem[{{Bla{\v z}ko}(1907)}]{Blazhko1907}
{Bla{\v z}ko}, S. 1907, Astronomische Nachrichten, 175, 325

\bibitem[{{Cacciari} {et~al.}(2005){Cacciari}, {Corwin}, \&
  {Carney}}]{Cacciari2005}
{Cacciari}, C., {Corwin}, T.~M., \& {Carney}, B.~W. 2005, \aj, 129, 267

\bibitem[{{Catelan}(2009)}]{Catelan2009}
{Catelan}, M. 2009, \apss, 320, 261

\bibitem[{{Catelan} {et~al.}(2004){Catelan}, {Pritzl}, \&
  {Smith}}]{Catelan2004}
{Catelan}, M., {Pritzl}, B.~J., \& {Smith}, H.~A. 2004, \apjs, 154, 633

\bibitem[{{Catelan} \& {Smith}(2015)}]{Catelan2015book}
{Catelan}, M. \& {Smith}, H.~A. 2015, {Pulsating Stars} (New York: Wiley)

\bibitem[{{Chadid} \& {Gillet}(1996{\natexlab{a}})}]{Chadid1996doub}
{Chadid}, M. \& {Gillet}, D. 1996{\natexlab{a}}, \aap, 308, 481

\bibitem[{{Chadid} \& {Gillet}(1996{\natexlab{b}})}]{Chadid1996broa}
{Chadid}, M. \& {Gillet}, D. 1996{\natexlab{b}}, \aap, 315, 475

\bibitem[{{Chadid} {et~al.}(2017){Chadid}, {Sneden}, \& {Preston}}]{Chadid2017}
{Chadid}, M., {Sneden}, C., \& {Preston}, G.~W. 2017, \apj, 835, 187

\bibitem[{{Chadid} {et~al.}(2014){Chadid}, {Vernin}, {Preston}, {Zalian},
  {Pouzenc}, {Abe}, {Agabi}, {Aristidi}, {Liu}, {M{\'e}karnia}, \&
  {Trinquet}}]{Chadid2014}
{Chadid}, M., {Vernin}, J., {Preston}, G., {et~al.} 2014, \aj, 148, 88

\bibitem[{{Christy}(1966)}]{Christy1966}
{Christy}, R.~F. 1966, \apj, 144, 108

\bibitem[{{Clementini} {et~al.}(2019){Clementini}, {Ripepi}, {Molinaro},
  {Garofalo}, {Muraveva}, {Rimoldini}, {Guy}, {Jevardat de Fombelle},
  {Nienartowicz}, {Marchal}, {Audard}, {Holl}, {Leccia}, {Marconi}, {Musella},
  {Mowlavi}, {Lecoeur-Taibi}, {Eyer}, {De Ridder}, {Regibo}, {Sarro},
  {Szabados}, {Evans}, \& {Riello}}]{Clementini2018}
{Clementini}, G., {Ripepi}, V., {Molinaro}, R., {et~al.} 2019, \aap, 622, A60

\bibitem[{{Cox}(1998)}]{Cox1998}
{Cox}, A.~N. 1998, \apj, 496, 246

\bibitem[{{D{\'e}k{\'a}ny} {et~al.}(2018){D{\'e}k{\'a}ny}, {Hajdu}, {Grebel},
  {Catelan}, {Elorrieta}, {Eyheramendy}, {Majaess}, \&
  {Jord{\'a}n}}]{Dekany2018}
{D{\'e}k{\'a}ny}, I., {Hajdu}, G., {Grebel}, E.~K., {et~al.} 2018, \apj, 857,
  54

\bibitem[{{D{\'e}k{\'a}ny} {et~al.}(2013){D{\'e}k{\'a}ny}, {Minniti},
  {Catelan}, {Zoccali}, {Saito}, {Hempel}, \& {Gonzalez}}]{Dekany2013}
{D{\'e}k{\'a}ny}, I., {Minniti}, D., {Catelan}, M., {et~al.} 2013, \apjl, 776,
  L19

\bibitem[{{Drake} {et~al.}(2013){Drake}, {Catelan}, {Djorgovski}, {Torrealba},
  {Graham}, {Belokurov}, {Koposov}, {Mahabal}, {Prieto}, {Donalek}, {Williams},
  {Larson}, {Christensen}, \& {Beshore}}]{Drake2013}
{Drake}, A.~J., {Catelan}, M., {Djorgovski}, S.~G., {et~al.} 2013, \apj, 763,
  32

\bibitem[{{Fabrizio} {et~al.}(2019){Fabrizio}, {Bono}, {Braga}, {Magurno},
  {Marinoni}, {Marrese}, {Ferraro}, {Fiorentino}, {Giuffrida}, {Iannicola},
  {Monelli}, {Altavilla}, {Chaboyer}, {Dall{\textquoteright}Ora}, {Gilligan},
  {Layden}, {Marengo}, {Nonino}, {Preston}, {Sesar}, {Sneden}, {Valenti},
  {Th{\'e}venin}, \& {Zoccali}}]{Fabrizio2019}
{Fabrizio}, M., {Bono}, G., {Braga}, V.~F., {et~al.} 2019, \apj, 882, 169

\bibitem[{{Fiorentino} {et~al.}(2015){Fiorentino}, {Bono}, {Monelli},
  {Stetson}, {Tolstoy}, {Gallart}, {Salaris}, {Mart{\'{\i}}nez-V{\'a}zquez}, \&
  {Bernard}}]{Fiorentino2015}
{Fiorentino}, G., {Bono}, G., {Monelli}, M., {et~al.} 2015, \apjl, 798, L12

\bibitem[{{Fokin}(1992)}]{Fokin1992}
{Fokin}, A.~B. 1992, \mnras, 256, 26

\bibitem[{{Gillet} \& {Crowe}(1988)}]{Gillet1988}
{Gillet}, D. \& {Crowe}, R.~A. 1988, \aap, 199, 242

\bibitem[{{Hajdu} {et~al.}(2015){Hajdu}, {Catelan}, {Jurcsik},
  {D{\'e}k{\'a}ny}, {Drake}, \& {Marquette}}]{Hajdu2015}
{Hajdu}, G., {Catelan}, M., {Jurcsik}, J., {et~al.} 2015, \mnras, 449, L113

\bibitem[{{Haschke} {et~al.}(2012{\natexlab{a}}){Haschke}, {Grebel}, \&
  {Duffau}}]{Haschke2012RRLyrLMC}
{Haschke}, R., {Grebel}, E.~K., \& {Duffau}, S. 2012{\natexlab{a}}, \aj, 144,
  106

\bibitem[{{Haschke} {et~al.}(2012{\natexlab{b}}){Haschke}, {Grebel}, \&
  {Duffau}}]{Haschke2012RRLyrSMC}
{Haschke}, R., {Grebel}, E.~K., \& {Duffau}, S. 2012{\natexlab{b}}, \aj, 144,
  107

\bibitem[{{Hernitschek} {et~al.}(2018){Hernitschek}, {Cohen}, {Rix}, {Sesar},
  {Martin}, {Magnier}, {Wainscoat}, {Kaiser}, {Tonry}, {Kudritzki}, {Hodapp},
  {Chambers}, {Flewelling}, \& {Burgett}}]{Hernitschek2018}
{Hernitschek}, N., {Cohen}, J.~G., {Rix}, H.-W., {et~al.} 2018, \apj, 859, 31

\bibitem[{{Hertzsprung}(1919)}]{Hertzsprung1919}
{Hertzsprung}, E. 1919, Astronomische Nachrichten, 210, 17

\bibitem[{{Hill}(1972)}]{Hill1972}
{Hill}, S.~J. 1972, \apj, 178, 793

\bibitem[{{Iglesias} \& {Rogers}(1996)}]{Iglesias1996}
{Iglesias}, C.~A. \& {Rogers}, F.~J. 1996, \apj, 464, 943

\bibitem[{{Jacyszyn-Dobrzeniecka} {et~al.}(2017){Jacyszyn-Dobrzeniecka},
  {Skowron}, {Mr{\'o}z}, {Soszy{\'n}ski}, {Udalski}, {Pietrukowicz}, {Skowron},
  {Poleski}, {Koz{\l}owski}, {Wyrzykowski}, {Pawlak}, {Szyma{\'n}ski}, \&
  {Ulaczyk}}]{JD2017RRlyrae}
{Jacyszyn-Dobrzeniecka}, A.~M., {Skowron}, D.~M., {Mr{\'o}z}, P., {et~al.}
  2017, \actaa, 67, 1

\bibitem[{{Jurcsik} {et~al.}(2001){Jurcsik}, {Clement}, {Geyer}, \&
  {Domsa}}]{Jurcsik2001}
{Jurcsik}, J., {Clement}, C., {Geyer}, E.~H., \& {Domsa}, I. 2001, \aj, 121,
  951

\bibitem[{{Jurcsik} {et~al.}(2012){Jurcsik}, {Hajdu}, {Szeidl}, {Ol{\'a}h},
  {Kelemen}, {S{\'o}dor}, {Saha}, {Mallick}, \& {Claver}}]{Jurcsik2012}
{Jurcsik}, J., {Hajdu}, G., {Szeidl}, B., {et~al.} 2012, \mnras, 419, 2173

\bibitem[{{Karczmarek} {et~al.}(2015){Karczmarek}, {Pietrzy{\'n}ski}, {Gieren},
  {Suchomska}, {Konorski}, {G{\'o}rski}, {Pilecki}, {Graczyk}, \&
  {Wielg{\'o}rski}}]{Karczmarek2015}
{Karczmarek}, P., {Pietrzy{\'n}ski}, G., {Gieren}, W., {et~al.} 2015, \aj, 150,
  90

\bibitem[{{Karczmarek} {et~al.}(2017){Karczmarek}, {Pietrzy{\'n}ski},
  {G{\'o}rski}, {Gieren}, \& {Bersier}}]{Karczmarek2017}
{Karczmarek}, P., {Pietrzy{\'n}ski}, G., {G{\'o}rski}, M., {Gieren}, W., \&
  {Bersier}, D. 2017, \aj, 154, 263

\bibitem[{{Koll{\'a}th} {et~al.}(2011){Koll{\'a}th}, {Moln{\'a}r}, \&
  {Szab{\'o}}}]{Kollath2011}
{Koll{\'a}th}, Z., {Moln{\'a}r}, L., \& {Szab{\'o}}, R. 2011, \mnras, 414, 1111

\bibitem[{{Koopmann} {et~al.}(1994){Koopmann}, {Lee}, {Demarque}, \&
  {Howard}}]{Koopmann1994}
{Koopmann}, R.~A., {Lee}, Y.-W., {Demarque}, P., \& {Howard}, J.~M. 1994, \apj,
  423, 380

\bibitem[{{Kunder} {et~al.}(2011){Kunder}, {Walker}, {Stetson}, {Bono},
  {Nemec}, {de Propris}, {Monelli}, {Cassisi}, {Andreuzzi}, {Dall'Ora}, {Di
  Cecco}, \& {Zoccali}}]{Kunder2011}
{Kunder}, A., {Walker}, A., {Stetson}, P.~B., {et~al.} 2011, \aj, 141, 15

\bibitem[{{Laskarides}(1974)}]{Laskarides1974}
{Laskarides}, P.~G. 1974, \apss, 27, 485

\bibitem[{{Le Borgne} {et~al.}(2007){Le Borgne}, {Paschke}, {Vandenbroere},
  {Poretti}, {Klotz}, {Bo{\"e}r}, {Damerdji}, {Martignoni}, \&
  {Acerbi}}]{LeBorgne2007}
{Le Borgne}, J.~F., {Paschke}, A., {Vandenbroere}, J., {et~al.} 2007, \aap,
  476, 307

\bibitem[{{Lee} {et~al.}(2014){Lee}, {L{\'o}pez-Morales}, {Hong}, {Kang},
  {Pohl}, \& {Walker}}]{Lee2014}
{Lee}, J.-W., {L{\'o}pez-Morales}, M., {Hong}, K., {et~al.} 2014, \apjs, 210, 6

\bibitem[{{Li{\v s}ka} {et~al.}(2016){Li{\v s}ka}, {Skarka}, {Zejda},
  {Mikul{\'a}{\v s}ek}, \& {de Villiers}}]{Liska2016b}
{Li{\v s}ka}, J., {Skarka}, M., {Zejda}, M., {Mikul{\'a}{\v s}ek}, Z., \& {de
  Villiers}, S.~N. 2016, \mnras, 459, 4360

\bibitem[{{Mateu} \& {Vivas}(2018)}]{Mateu2018}
{Mateu}, C. \& {Vivas}, A.~K. 2018, \mnras, 479, 211

\bibitem[{{Minniti} {et~al.}(2010){Minniti}, {Lucas}, {Emerson}, {Saito},
  {Hempel}, {Pietrukowicz}, {Ahumada}, {Alonso}, {Alonso-Garcia}, {Arias},
  {Bandyopadhyay}, {Barb{\'a}}, {Barbuy}, {Bedin}, {Bica}, {Borissova},
  {Bronfman}, {Carraro}, {Catelan}, {Clari{\'a}}, {Cross}, {de Grijs},
  {D{\'e}k{\'a}ny}, {Drew}, {Fari{\~n}a}, {Feinstein}, {Fern{\'a}ndez
  Laj{\'u}s}, {Gamen}, {Geisler}, {Gieren}, {Goldman}, {Gonzalez}, {Gunthardt},
  {Gurovich}, {Hambly}, {Irwin}, {Ivanov}, {Jord{\'a}n}, {Kerins}, {Kinemuchi},
  {Kurtev}, {L{\'o}pez-Corredoira}, {Maccarone}, {Masetti}, {Merlo},
  {Messineo}, {Mirabel}, {Monaco}, {Morelli}, {Padilla}, {Palma}, {Parisi},
  {Pignata}, {Rejkuba}, {Roman-Lopes}, {Sale}, {Schreiber}, {Schr{\"o}der},
  {Smith}, {}, {Soto}, {Tamura}, {Tappert}, {Thompson}, {Toledo}, {Zoccali}, \&
  {Pietrzynski}}]{Minniti2010VVV}
{Minniti}, D., {Lucas}, P.~W., {Emerson}, J.~P., {et~al.} 2010, \na, 15, 433

\bibitem[{{Muraveva} {et~al.}(2018){Muraveva}, {Delgado}, {Clementini},
  {Sarro}, \& {Garofalo}}]{Muraveva2018}
{Muraveva}, T., {Delgado}, H.~E., {Clementini}, G., {Sarro}, L.~M., \&
  {Garofalo}, A. 2018, \mnras, 481, 1195

\bibitem[{{Oosterhoff}(1939)}]{Oosterhoff1939}
{Oosterhoff}, P.~T. 1939, The Observatory, 62, 104

\bibitem[{{Paparo} {et~al.}(1998){Paparo}, {Saad}, {Szeidl}, {Kollath}, {Abu
  Elazm}, \& {Sharaf}}]{Paparo1998}
{Paparo}, M., {Saad}, S.~M., {Szeidl}, B., {et~al.} 1998, \aap, 332, 102

\bibitem[{{Pietrukowicz} {et~al.}(2015){Pietrukowicz}, {Koz{\l}owski},
  {Skowron}, {Soszy{\'n}ski}, {Udalski}, {Poleski}, {Wyrzykowski},
  {Szyma{\'n}ski}, {Pietrzy{\'n}ski}, {Ulaczyk}, {Mr{\'o}z}, {Skowron}, \&
  {Kubiak}}]{Pietrukowicz2015}
{Pietrukowicz}, P., {Koz{\l}owski}, S., {Skowron}, J., {et~al.} 2015, \apj,
  811, 113

\bibitem[{{Pietrzy{\'n}ski} {et~al.}(2008){Pietrzy{\'n}ski}, {Gieren},
  {Szewczyk}, {Walker}, {Rizzi}, {Bresolin}, {Kudritzki}, {Nalewajko}, {Storm},
  {Dall'Ora}, \& {Ivanov}}]{Pietrzynski2008}
{Pietrzy{\'n}ski}, G., {Gieren}, W., {Szewczyk}, O., {et~al.} 2008, \aj, 135,
  1993

\bibitem[{{Preston}(2009)}]{Preston2009}
{Preston}, G.~W. 2009, \aap, 507, 1621

\bibitem[{{Preston} \& {Paczynski}(1964)}]{Preston1964}
{Preston}, G.~W. \& {Paczynski}, B. 1964, \apj, 140, 181

\bibitem[{{Preston} {et~al.}(1965){Preston}, {Smak}, \&
  {Paczynski}}]{Preston1965}
{Preston}, G.~W., {Smak}, J., \& {Paczynski}, B. 1965, \apjs, 12, 99

\bibitem[{{Prudil} {et~al.}(2019{\natexlab{a}}){Prudil}, {D{\'e}k{\'a}ny},
  {Catelan}, {Smolec}, {Grebel}, \& {Skarka}}]{Prudil2019OO}
{Prudil}, Z., {D{\'e}k{\'a}ny}, I., {Catelan}, M., {et~al.} 2019{\natexlab{a}},
  \mnras, 484, 4833

\bibitem[{{Prudil} \& {Skarka}(2017)}]{Prudil2017Blazhko}
{Prudil}, Z. \& {Skarka}, M. 2017, \mnras, 466, 2602

\bibitem[{{Prudil} {et~al.}(2019{\natexlab{b}}){Prudil}, {Skarka},
  {Li{\v{s}}ka}, {Grebel}, \& {Lee}}]{Prudil2019Binary}
{Prudil}, Z., {Skarka}, M., {Li{\v{s}}ka}, J., {Grebel}, E.~K., \& {Lee}, C.~U.
  2019{\natexlab{b}}, \mnras, 487, L1

\bibitem[{{Sandage}(1981)}]{Sandage1981}
{Sandage}, A. 1981, \apj, 248, 161

\bibitem[{{Sandage} {et~al.}(1981){Sandage}, {Katem}, \&
  {Sandage}}]{Sandage1981KS}
{Sandage}, A., {Katem}, B., \& {Sandage}, M. 1981, \apjs, 46, 41

\bibitem[{{Silva Aguirre} {et~al.}(2008){Silva Aguirre}, {Catelan}, {Weiss}, \&
  {Valcarce}}]{Silva-Aguirre2008}
{Silva Aguirre}, V., {Catelan}, M., {Weiss}, A., \& {Valcarce}, A.~A.~R. 2008,
  \aap, 489, 1201

\bibitem[{{Smolec} \& {Moskalik}(2008)}]{Smolec2008}
{Smolec}, R. \& {Moskalik}, P. 2008, \actaa, 58, 193

\bibitem[{{Smolec} {et~al.}(2015){Smolec}, {Soszy{\'n}ski}, {Udalski},
  {Szyma{\'n}ski}, {Pietrukowicz}, {Skowron}, {Koz{\l}owski}, {Poleski},
  {Moskalik}, {Skowron}, {Pietrzy{\'n}ski}, {Wyrzykowski}, {Ulaczyk}, \&
  {Mr{\'o}z}}]{Smolec2015}
{Smolec}, R., {Soszy{\'n}ski}, I., {Udalski}, A., {et~al.} 2015, \mnras, 447,
  3873

\bibitem[{{Soszy{\'n}ski} {et~al.}(2011){Soszy{\'n}ski}, {Dziembowski},
  {Udalski}, {Poleski}, {Szyma{\'n}ski}, {Kubiak}, {Pietrzy{\'n}ski},
  {Wyrzykowski}, {Ulaczyk}, {Koz{\l}owski}, \&
  {Pietrukowicz}}]{Soszynski2011OGLEIII}
{Soszy{\'n}ski}, I., {Dziembowski}, W.~A., {Udalski}, A., {et~al.} 2011,
  \actaa, 61, 1

\bibitem[{{Soszy{\'n}ski} {et~al.}(2014){Soszy{\'n}ski}, {Udalski},
  {Szyma{\'n}ski}, {Pietrukowicz}, {Mr{\'o}z}, {Skowron}, {Koz{\l}owski},
  {Poleski}, {Skowron}, {Pietrzy{\'n}ski}, {Wyrzykowski}, {Ulaczyk}, \&
  {Kubiak}}]{Soszynski2014OGLEIV}
{Soszy{\'n}ski}, I., {Udalski}, A., {Szyma{\'n}ski}, M.~K., {et~al.} 2014,
  \actaa, 64, 177

\bibitem[{{Soszy{\'n}ski} {et~al.}(2017){Soszy{\'n}ski}, {Udalski},
  {Szyma{\'n}ski}, {Wyrzykowski}, {Ulaczyk}, {Poleski}, {Pietrukowicz},
  {Koz{\l}owski}, {Skowron}, {Skowron}, {Mr{\'o}z}, {Pawlak}, {Rybicki}, \&
  {Jacyszyn-Dobrzeniecka}}]{Soszynski2017OGLEIV}
{Soszy{\'n}ski}, I., {Udalski}, A., {Szyma{\'n}ski}, M.~K., {et~al.} 2017,
  \actaa, 67, 297

\bibitem[{{Struve} \& {Blaauw}(1948)}]{Struve1948}
{Struve}, O. \& {Blaauw}, A. 1948, \apj, 108, 60

\bibitem[{{Sweigart} \& {Renzini}(1979)}]{Sweigart1979}
{Sweigart}, A.~V. \& {Renzini}, A. 1979, \aap, 71, 66

\bibitem[{{Szab{\'o}} {et~al.}(2010){Szab{\'o}}, {Koll{\'a}th}, {Moln{\'a}r},
  {Kolenberg}, {Kurtz}, {Bryson}, {Benk{\H o}}, {Christensen-Dalsgaard},
  {Kjeldsen}, {Borucki}, {Koch}, {Twicken}, {Chadid}, {di Criscienzo}, {Jeon},
  {Moskalik}, {Nemec}, \& {Nuspl}}]{Szabo2010}
{Szab{\'o}}, R., {Koll{\'a}th}, Z., {Moln{\'a}r}, L., {et~al.} 2010, \mnras,
  409, 1244

\bibitem[{{Udalski} {et~al.}(1994){Udalski}, {Kubiak}, {Szymanski}, {Kaluzny},
  {Mateo}, \& {Krzeminski}}]{Udalski1994I}
{Udalski}, A., {Kubiak}, M., {Szymanski}, M., {et~al.} 1994, \actaa, 44, 317

\bibitem[{{Udalski} {et~al.}(1996){Udalski}, {Olech}, {Szymanski}, {Kaluzny},
  {Kubiak}, {Krzeminski}, {Mateo}, \& {Stanek}}]{Udalski1996IV}
{Udalski}, A., {Olech}, A., {Szymanski}, M., {et~al.} 1996, \actaa, 46, 51

\bibitem[{{Udalski} {et~al.}(1995{\natexlab{a}}){Udalski}, {Olech},
  {Szymanski}, {Kaluzny}, {Kubiak}, {Mateo}, \& {Krzeminski}}]{Udalski1995III}
{Udalski}, A., {Olech}, A., {Szymanski}, M., {et~al.} 1995{\natexlab{a}},
  \actaa, 45, 433

\bibitem[{{Udalski} {et~al.}(1997){Udalski}, {Olech}, {Szymanski}, {Kaluzny},
  {Kubiak}, {Mateo}, {Krzeminski}, \& {Stanek}}]{Udalski1997V}
{Udalski}, A., {Olech}, A., {Szymanski}, M., {et~al.} 1997, \actaa, 47, 1

\bibitem[{{Udalski} {et~al.}(1995{\natexlab{b}}){Udalski}, {Szymanski},
  {Kaluzny}, {Kubiak}, {Mateo}, \& {Krzeminski}}]{Udalski1995II}
{Udalski}, A., {Szymanski}, M., {Kaluzny}, J., {et~al.} 1995{\natexlab{b}},
  \actaa, 45, 1

\bibitem[{{Udalski} {et~al.}(2015){Udalski}, {Szyma{\'n}ski}, \&
  {Szyma{\'n}ski}}]{Udalski2015}
{Udalski}, A., {Szyma{\'n}ski}, M.~K., \& {Szyma{\'n}ski}, G. 2015, \actaa, 65,
  1

\bibitem[{{van Albada} \& {Baker}(1973)}]{vanAlbada1973}
{van Albada}, T.~S. \& {Baker}, N. 1973, \apj, 185, 477

\end{thebibliography}

\begin{appendix}
\section{Table with calculated properties of studied RR~Lyrae stars} \label{sec:AppTable}
\begin{landscape}
\begin{table}
\tiny
\caption{The first lines of a table for the photometric and shock properties of the studied RR~Lyrae stars. The full table can be found in the supplementary material. The first column contains identification names for individual stars in the form of OGLE-BLG-RRLYR-ID. Columns 2 and 3 list the mean apparent magnitudes and amplitudes, respectively, in the $K_{s}$-band from the VVV photometric survey \citep{Minniti2010VVV}. Columns 4 and 5 contain metallicity estimates and their errors calculated from the $I$ band light curves. Columns 6, 7, 8, 9, 10, and 11 contain absolute magnitudes and their errors for the studied stars in the $V$, $I$, and $K_{s}$-bands, respectively. The approximate centers of the humps and bumps are listed in columns 12, 13, 14, 15. Columns 16 and 17 contain the determined sizes of both shocks in the phased light curves. Column 18 lists the association with the Oosterhoff groups (1 -- Oo\,I, 2 -- Oo\,II). Columns 19 and 20 contain the parameter $a_{2}$ and its error used to estimate the period change rate of the studied stars. The last two columns 21 and 22 list the categories 1 -- 4 from the visual classification of the humps and bumps.}
\label{tab:KinProp}
\setlength\tabcolsep{3pt}
\begin{tabular}{lccccccccccccccccccccc}
\hline \hline
ID                & $K_{s}$                  & Amp$^{K_{s}}$                 & [Fe/H]                & $\sigma_{\rm [Fe/H]}$          & $M_{V}$                & $\sigma_{M_{V}}$            & $M_{I}$                 & $\sigma_{M_{I}}$            & $M_{K_{s}}$                 & $\sigma_{M_{K_{s}}}$            & $x_{\rm HUMP}$            & $y_{\rm HUMP}$            & $x_{\rm BUMP}$            & $y_{\rm BUMP}$            & Amp$_{\rm HUMP}$ & Amp$_{\rm BUMP}$            & Oo    & $a_{2}$ & $\sigma_{a_{2}}$    & Class$_{\rm HUMP}$ & Class$_{\rm BUMP}$ \\
 & [mag] & [mag] & [dex] & [dex] & [mag] & [mag] & [mag] & [mag] & [mag] & [mag] & & & & & & & & & & &\\ \hline
01010 & 13.014 & 0.095 & -1.265 & 0.244 & 0.605 & 0.230 & 0.214 & 0.050 & -0.277 & 0.041 & -0.064 & -0.1067 & -0.3038 & 0.2360 & 0.0827 & 0.0510 & 1 & 2.463$\cdot$\,10$^{-10}$ & 7.945$\cdot$\,10$^{-12}$ & 2 & 2 \\ 
01058 & 13.779 & 0.313 & -1.004 & 0.171 & 0.672 & 0.159 & 0.209 & 0.035 & -0.363 & 0.029 & -0.062 & -0.1299 & -0.2828 & 0.2614 & 0.0897 & 0.0710 & 1 & -2.943$\cdot$\,10$^{-10}$ & 1.897$\cdot$\,10$^{-11}$ & 2 & 3 \\ 
01082 & 13.659 & 0.222 & -1.143 & 0.207 & 0.635 & 0.194 & 0.092 & 0.042 & -0.590 & 0.035 & -0.111 & -0.1757 & -0.3833 & 0.1762 & 0.1602 & 0.0521 & 1 & 3.074$\cdot$\,10$^{-12}$ & 9.346$\cdot$\,10$^{-12}$ & 3 & 2 \\ 
01646 & 13.693 & 0.198 & -1.023 & 0.243 & 0.667 & 0.227 & 0.120 & 0.050 & -0.561 & 0.041 & -0.131 & -0.1644 & -0.4238 & 0.1852 & 0.0440 & 0.0287 & 1 & -1.309$\cdot$\,10$^{-11}$ & 4.212$\cdot$\,10$^{-11}$ & 1 & 1 \\ 
01692 & 14.200 & 0.359 & -1.039 & 0.197 & 0.662 & 0.184 & 0.167 & 0.040 & -0.449 & 0.033 & -0.067 & -0.1016 & -0.3021 & 0.2687 & 0.0466 & 0.0671 & 2 & -1.586$\cdot$\,10$^{-10}$ & 9.942$\cdot$\,10$^{-12}$ & 1 & 4 \\ 
01694 & 14.111 & 0.279 & -1.217 & 0.194 & 0.617 & 0.182 & 0.069 & 0.040 & -0.620 & 0.033 & -0.073 & -0.1019 & -0.3808 & 0.1946 & 0.0740 & 0.0484 & 2 & -4.227$\cdot$\,10$^{-10}$ & 6.114$\cdot$\,10$^{-11}$ & 2 & 3 \\ 
01732 & 14.641 & 0.388 & -0.883 & 0.211 & 0.708 & 0.196 & 0.372 & 0.043 & -0.027 & 0.036 & -0.063 & -0.0279 & -0.2476 & 0.3355 & 0.0158 & 0.0416 & 1 & 8.722$\cdot$\,10$^{-11}$ & 3.240$\cdot$\,10$^{-12}$ & 1 & 3 \\ 
01817 & 12.917 & 0.070 & -2.113 & 0.310 & 0.492 & 0.303 & 0.016 & 0.064 & -0.480 & 0.053 & -0.059 & -0.0256 & -0.2526 & 0.3286 & 0.0336 & 0.0427 & 1 & -1.138$\cdot$\,10$^{-10}$ & 3.731$\cdot$\,10$^{-11}$ & 1 & 3 \\ 
01954 & 13.864 & 0.282 & -1.507 & 0.194 & 0.557 & 0.184 & -0.002 & 0.040 & -0.699 & 0.033 & -0.074 & -0.0589 & -0.3472 & 0.2485 & 0.0481 & 0.0547 & 2 & 2.991$\cdot$\,10$^{-10}$ & 9.175$\cdot$\,10$^{-11}$ & 1 & 3 \\ 
01959 & 14.222 & 0.203 & -1.018 & 0.221 & 0.668 & 0.206 & 0.092 & 0.045 & -0.627 & 0.038 & -0.124 & -0.1782 & -0.3704 & 0.2177 & 0.1467 & 0.0423 & 2 & 6.636$\cdot$\,10$^{-11}$ & 2.444$\cdot$\,10$^{-11}$ & 3 & 1 \\ 
02019 & 14.226 & 0.272 & -1.010 & 0.198 & 0.670 & 0.184 & 0.157 & 0.041 & -0.482 & 0.034 & -0.110 & -0.1805 & -0.3719 & 0.1989 & 0.1409 & 0.0698 & 1 & -2.651$\cdot$\,10$^{-11}$ & 2.995$\cdot$\,10$^{-11}$ & 3 & 2 \\ 
02050 & 14.455 & 0.264 & -1.120 & 0.173 & 0.641 & 0.162 & 0.202 & 0.035 & -0.346 & 0.029 & -0.064 & -0.0948 & -0.2681 & 0.2848 & 0.0595 & 0.0674 & 1 & -6.851$\cdot$\,10$^{-11}$ & 1.861$\cdot$\,10$^{-11}$ & 2 & 3 \\ 
02060 & 14.173 & 0.332 & -1.400 & 0.178 & 0.577 & 0.168 & 0.048 & 0.036 & -0.616 & 0.030 & -0.079 & -0.1299 & -0.2956 & 0.2422 & 0.1287 & 0.0856 & 2 & 2.627$\cdot$\,10$^{-12}$ & 2.042$\cdot$\,10$^{-11}$ & 2 & 3 \\ 
02062 & 14.183 & 0.317 & -1.489 & 0.175 & 0.560 & 0.166 & 0.041 & 0.036 & -0.605 & 0.030 & -0.065 & -0.0970 & -0.2774 & 0.2851 & 0.0386 & 0.0737 & 2 & -3.015$\cdot$\,10$^{-10}$ & 7.012$\cdot$\,10$^{-11}$ & 2 & 4 \\ 
02083 & 14.309 & 0.191 & -1.022 & 0.235 & 0.667 & 0.219 & 0.141 & 0.048 & -0.513 & 0.040 & -0.141 & -0.1276 & -0.3976 & 0.2195 & 0.0972 & 0.0286 & 1 & 2.851$\cdot$\,10$^{-11}$ & 4.972$\cdot$\,10$^{-11}$ & 1 & 1 \\ 
02105 & 13.838 & 0.350 & -1.264 & 0.203 & 0.606 & 0.191 & 0.089 & 0.042 & -0.560 & 0.035 & -0.072 & -0.0718 & -0.3360 & 0.2523 & 0.0354 & 0.0553 & 2 & 1.438$\cdot$\,10$^{-10}$ & 3.503$\cdot$\,10$^{-11}$ & 1 & 4 \\ 
02138 & 14.174 & 0.231 & -1.105 & 0.229 & 0.644 & 0.214 & 0.051 & 0.047 & -0.694 & 0.039 & -0.125 & -0.1782 & -0.3870 & 0.1968 & 0.1681 & 0.0388 & 2 & 9.229$\cdot$\,10$^{-11}$ & 2.954$\cdot$\,10$^{-11}$ & 4 & 1 \\ 
02150 & 14.537 & 0.139 & -1.183 & 0.262 & 0.625 & 0.246 & 0.048 & 0.054 & -0.679 & 0.045 & -0.158 & -0.0809 & -0.4253 & 0.2269 & 0.0460 & 0.0134 & 1 & -4.280$\cdot$\,10$^{-10}$ & 3.396$\cdot$\,10$^{-10}$ & 1 & 1 \\ 
02185 & 14.162 & 0.190 & -1.131 & 0.219 & 0.638 & 0.205 & 0.082 & 0.045 & -0.617 & 0.037 & -0.126 & -0.1740 & -0.3871 & 0.1824 & 0.1262 & 0.0407 & 1 & 1.214$\cdot$\,10$^{-10}$ & 2.147$\cdot$\,10$^{-10}$ & 3 & 1 \\ 
02218 & 14.554 & 0.400 & -1.042 & 0.201 & 0.661 & 0.188 & 0.287 & 0.041 & -0.176 & 0.034 & -0.061 & -0.0300 & -0.2219 & 0.3634 & 0.0239 & 0.0509 & 1 & -1.232$\cdot$\,10$^{-10}$ & 1.114$\cdot$\,10$^{-11}$ & 1 & 3 \\ 
02238 & 13.952 & 0.276 & -1.542 & 0.239 & 0.551 & 0.227 & -0.096 & 0.049 & -0.902 & 0.041 & -0.091 & -0.1551 & -0.3564 & 0.2361 & 0.1206 & 0.0322 & 2 & -1.937$\cdot$\,10$^{-11}$ & 7.667$\cdot$\,10$^{-11}$ & 2 & 1 \\ 
02260 & 14.425 & 0.335 & -1.052 & 0.177 & 0.659 & 0.165 & 0.230 & 0.036 & -0.302 & 0.030 & -0.066 & -0.0387 & -0.2552 & 0.3064 & 0.0373 & 0.0621 & 1 & -1.422$\cdot$\,10$^{-10}$ & 2.779$\cdot$\,10$^{-11}$ & 1 & 3 \\ 
02288 & 14.330 & 0.304 & -1.086 & 0.170 & 0.649 & 0.159 & 0.171 & 0.035 & -0.426 & 0.029 & -0.070 & -0.1274 & -0.2916 & 0.2523 & 0.1003 & 0.0719 & 1 & -1.222$\cdot$\,10$^{-10}$ & 2.383$\cdot$\,10$^{-11}$ & 2 & 3 \\ 
02290 & 14.397 & 0.253 & -1.055 & 0.211 & 0.658 & 0.197 & 0.126 & 0.043 & -0.539 & 0.036 & -0.119 & -0.1581 & -0.3639 & 0.2092 & 0.1358 & 0.0518 & 1 & 7.020$\cdot$\,10$^{-11}$ & 1.902$\cdot$\,10$^{-11}$ & 3 & 2 \\ 
02357 & 13.774 & 0.285 & -1.486 & 0.193 & 0.561 & 0.183 & -0.013 & 0.040 & -0.730 & 0.033 & -0.084 & -0.1392 & -0.3261 & 0.2433 & 0.1410 & 0.0644 & 2 & -6.056$\cdot$\,10$^{-10}$ & 1.342$\cdot$\,10$^{-10}$ & 2 & 3 \\ 
02418 & 14.378 & 0.254 & -1.088 & 0.177 & 0.649 & 0.165 & 0.143 & 0.036 & -0.489 & 0.030 & -0.079 & -0.1621 & -0.3217 & 0.2221 & 0.1428 & 0.0627 & 1 & 3.791$\cdot$\,10$^{-12}$ & 1.494$\cdot$\,10$^{-11}$ & 2 & 3 \\ 
02590 & 14.259 & 0.298 & -1.105 & 0.170 & 0.644 & 0.159 & 0.196 & 0.035 & -0.363 & 0.029 & -0.064 & -0.1247 & -0.2811 & 0.2661 & 0.0930 & 0.0719 & 1 & 2.301$\cdot$\,10$^{-9}$ & 3.606$\cdot$\,10$^{-10}$ & 2 & 4 \\ 
02687 & 14.693 & 0.289 & -1.125 & 0.173 & 0.639 & 0.162 & 0.176 & 0.035 & -0.404 & 0.029 & -0.073 & -0.1481 & -0.2995 & 0.2384 & 0.1325 & 0.0832 & 1 & 2.037$\cdot$\,10$^{-10}$ & 1.022$\cdot$\,10$^{-11}$ & 2 & 3 \\ 
02814 & 14.168 & 0.252 & -1.104 & 0.191 & 0.645 & 0.179 & 0.120 & 0.039 & -0.538 & 0.032 & -0.100 & -0.1645 & -0.3501 & 0.2269 & 0.1570 & 0.0644 & 1 & 5.179$\cdot$\,10$^{-11}$ & 2.704$\cdot$\,10$^{-11}$ & 3 & 2 \\ 
\dots & \dots & \dots & \dots & \dots & \dots & \dots & \dots & \dots & \dots & \dots & \dots & \dots & \dots & \dots & \dots & \dots & \dots & \dots & \dots & \dots & \dots \\
\hline
\end{tabular}
\end{table}
\end{landscape}  
\end{appendix}

\end{document}